\documentclass[fleqn,usenatbib]{mnras}

\usepackage{newtxtext,newtxmath}

\usepackage[T1]{fontenc}
\usepackage{ae,aecompl}

\usepackage{graphicx}	
\usepackage{amsmath}	
\usepackage{siunitx}
\usepackage{booktabs}
\usepackage{multirow}
\usepackage{color, colortbl}
\usepackage[pdftex,dvipsnames]{xcolor}
\usepackage[ruled,vlined]{algorithm2e}
\usepackage{lineno}
\usepackage{enumitem}
\usepackage[makeroom]{cancel}

\newcommand{\buzzard}{\textsc{Buzzard}\xspace}
\newcommand{\sompz}{\textsc{SOMPZ}\xspace}
\newcommand{\balrog}{\textsc{Balrog}\xspace}

\newcommand{\pz}{$p(z)$\xspace}
\newcommand{\pzi}{$p_i(z)$\xspace}
\newcommand{\pzfid}{$p^{\mathrm{fid.}}(z)$\xspace}

\newcommand{\nz}{$n(z)$\xspace}

\newcommand{\nzfid}{$n^{\mathrm{fid.}}(z)$\xspace}

\newcommand{\pzA}{$p^{\mathrm{in.}}(z)$\xspace}
\newcommand{\pzAi}{$p_i^{\mathrm{in.}}(z)$\xspace}
\newcommand{\pzAavg}{$p^{\langle \mathrm{in.} \rangle}(z)$\xspace}
\newcommand{\pzB}{$p^{\mathrm{out.}}(z)$\xspace}
\newcommand{\pzBi}{$p_i^{\mathrm{out.}}(z)$\xspace}

\newcommand{\nzA}{$n^{\mathrm{in.}}(z)$\xspace}

\newcommand{\nzB}{$n^{\mathrm{out.}}(z)$\xspace}

\newcommand{\sigmacritinvmm}{\Sigma^{-1}_{\mathrm{crit.}}}
\newcommand{\avgsigmacritinv}{$\overline{ \Sigma^{-1}}_{\mathrm{crit.}}$\xspace}
\newcommand{\avgsigmacritinvmm}{\overline{ \Sigma^{-1}}_{\mathrm{crit.}}}
\newcommand{\avgsigmacritinvB}{$\overline{ \Sigma^{-1, \mathrm{out.}}}_{\mathrm{crit.}}$\xspace}

\usepackage{xcolor}
\usepackage{ulem}
\usepackage{xspace} 
\usepackage{hyperref}

\usepackage{threeparttable}
\usepackage{eso-pic}

\AddToShipoutPictureBG*{%
  \AtPageUpperLeft{%
    \hspace{0.75\paperwidth}%
    \raisebox{-3.5\baselineskip}{%
      \makebox[0pt][l]{\textnormal{DES-2022-0692}}
}}}%

\AddToShipoutPictureBG*{%
  \AtPageUpperLeft{%
    \hspace{0.75\paperwidth}%
    \raisebox{-4.5\baselineskip}{%
      \makebox[0pt][l]{\textnormal{FERMILAB-PUB-22-295-PPD}}
}}}%

\title[PITPZ]{Mapping Variations of Redshift Distributions with Probability Integral Transforms}

\author[J. Myles et al.]{
\parbox{\textwidth}{
\Large
J.~Myles,$^{1,2,3,4}$
D.~Gruen,$^{4}$
A.~Amon,$^{5,6}$
A.~Alarcon,$^{7}$
J.~DeRose,$^{8}$
S.~Everett,$^{9}$
S.~Dodelson,$^{10,11}$
G.~M.~Bernstein,$^{12}$
A.~Campos,$^{10}$
I.~Harrison,$^{13}$
N.~MacCrann,$^{14}$
J.~McCullough,$^{2}$
M.~Raveri,$^{12}$
C.~S{\'a}nchez,$^{12}$
M.~A.~Troxel,$^{15}$
B.~Yin,$^{10}$
T.~M.~C.~Abbott,$^{16}$
S.~Allam,$^{17}$
O.~Alves,$^{18}$
F.~Andrade-Oliveira,$^{18}$
E.~Bertin,$^{19,20}$
D.~Brooks,$^{21}$
D.~L.~Burke,$^{2,3}$
A.~Carnero~Rosell,$^{22,23,24}$
M.~Carrasco~Kind,$^{25,26}$
J.~Carretero,$^{27}$
R.~Cawthon,$^{28}$
M.~Costanzi,$^{29,30,31}$
L.~N.~da Costa,$^{23}$
M.~E.~S.~Pereira,$^{32}$
S.~Desai,$^{33}$
P.~Doel,$^{21}$
I.~Ferrero,$^{34}$
B.~Flaugher,$^{17}$
J.~Frieman,$^{17,35}$
J.~Garc\'ia-Bellido,$^{36}$
M.~Gatti,$^{12}$
D.~W.~Gerdes,$^{37,18}$
R.~A.~Gruendl,$^{25,26}$
J.~Gschwend,$^{23,38}$
G.~Gutierrez,$^{17}$
W.~G.~Hartley,$^{39}$
S.~R.~Hinton,$^{40}$
D.~L.~Hollowood,$^{41}$
K.~Honscheid,$^{42,43}$
D.~J.~James,$^{44}$
K.~Kuehn,$^{45,46}$
O.~Lahav,$^{21}$
P.~Melchior,$^{47}$
J. Mena-Fern{\'a}ndez,$^{48}$
F.~Menanteau,$^{25,26}$
R.~Miquel,$^{49,27}$
J.~J.~Mohr,$^{50,4}$
A.~Palmese,$^{51}$
F.~Paz-Chinch\'{o}n,$^{25,5}$
A.~Pieres,$^{23,38}$
A.~A.~Plazas~Malag\'on,$^{47}$
J.~Prat,$^{52,35}$
M.~Rodriguez-Monroy,$^{48}$
E.~Sanchez,$^{48}$
V.~Scarpine,$^{17}$
I.~Sevilla-Noarbe,$^{48}$
M.~Smith,$^{53}$
E.~Suchyta,$^{54}$
M.~E.~C.~Swanson,
G.~Tarle,$^{18}$
D.~L.~Tucker,$^{17}$
M.~Vincenzi,$^{55,53}$
and N.~Weaverdyck$^{18,8}$
\begin{center} (DES Collaboration) \end{center}
}
\vspace{0.2cm}
\\
\parbox{\textwidth}{ \small
\textit{The authors' affiliations are shown at the end of this paper.}}}

\date{Accepted 2022 December 2. Received 2022 November 21; in original form 2022 August 24}

\pubyear{2022}
\begin{document}
\label{firstpage}
\pagerange{\pageref{firstpage}--\pageref{lastpage}}
\maketitle

\begin{abstract}
We present a method for mapping variations between probability distribution functions and apply this method within the context of measuring galaxy redshift distributions from imaging survey data. This method, which we name PITPZ for the probability integral transformations it relies on, uses a difference in curves between distribution functions in an ensemble as a transformation to apply to another distribution function, thus transferring the variation in the ensemble to the latter distribution function. This procedure is broadly applicable to the problem of uncertainty propagation. In the context of redshift distributions, for example, the uncertainty contribution due to certain effects can be studied effectively only in simulations, thus necessitating a transfer of variation measured in simulations to the redshift distributions measured from data. We illustrate the use of PITPZ by using the method to propagate photometric calibration uncertainty to redshift distributions of the Dark Energy Survey Year 3 weak lensing source galaxies. For this test case, we find that PITPZ yields a lensing amplitude uncertainty estimate due to photometric calibration error within 1 per cent of the truth, compared to as much as a 30 per cent underestimate when using traditional methods. 
\end{abstract}




\section{Introduction} 
\label{sec:intro}

The matter density field of the Universe and its evolution over time relate directly to the cosmological model of the Universe. Galaxy surveys provide observable proxies of the matter density field and thus can be used to place competitive constraints on parameters of cosmological models. Specifically, experiments such as the Dark Energy Survey (DES), Kilo-Degree Survey (KiDS), and the Hyper Suprime-Cam Survey (HSC) as well as the future Vera C. Rubin Observatory's Legacy Survey of Space and Time (LSST), \textit{Euclid}, and \textit{Roman} Space Telescope missions measure statistics such as correlation functions of galaxy positions and shapes to probe the underlying matter density field \citep{euclid, roman, lsst_desc, kids450, desy1_3x2pt, HSC_hikage_19, kidsviking450, kids1000_3x2pt,  y3-3x2ptkp}. In these analyses, determining the impact of weak gravitational lensing on the observed galaxy images provides crucial information to relate observations to the underlying matter density field that galaxies live in. Among the data products needed for these experiments, \textit{redshift distributions}, which encode the relative contribution of galaxies at different redshifts to the gravitational lensing signal observed, loom large due to their key role in enabling interpretation of the effect of weak lensing on the apparent shapes and sizes of galaxies \citetext{For a review, see e.g. \citealt{NewmanGruen2022}. See also \citealp{Huterer2006,Lima2008, cfht_hilde, cunha2012, Benjamin2013,Huterer2013,Bonnett2016, Samuroff2017};
\citealp*{Hoyle2018};
\citealp{Wright2020a, Wright2020b, KidsDEScomb, Tessore2020, kids1000_nz, Euclid2020};
\citealp*{Myles2021, y3-sourcewz, y3-shearratio};
\citealp{Cabayol2022}
}. 

In lensing survey nomenclature, the term `redshift distribution' refers to a function describing the relative probability of a galaxy in a sample to have come from a particular narrow redshift histogram bin. A typical lensing survey will divide its dataset into a few tomographic bins, each with its own redshift distribution. We highlight that a redshift distribution is distinct from the photometric redshift for any individual galaxy, and the uncertainty requirements of redshift distributions are likewise distinct from uncertainty requirements of individual galaxy photometric redshifts. As reducing systematic uncertainties in redshift distributions is necessary to meet uncertainty goals on estimated cosmological parameters, greater attention is being drawn to the importance of modelling redshift distribution uncertainty with sufficient complexity (see e.g. \citealt*{Myles2021}; \citealt{Malz2018, Hadzhiyska2020, Malz2021, Stolzner2021, y3-hyperrank, Zhang2022,  Malz2022}). Redshift distributions have been historically described as a single probability density function together with, for example, a shift parameter describing uncertainty on the mean redshift value (e.g. \citealt{Hoyle2018}). More recently, redshift distributions have been described as \textit{joint} probability distribution function (PDF) for redshift histogram bin heights, meaning each bin in a redshift histogram has a full associated PDF (see e.g. \citealt{Leistedt2019, Sanchez2019, Alarcon2020}) or alternatively as an \textit{ensemble} of slightly varying PDFs that collectively describe the full uncertainty in knowledge of galaxy redshift (see e.g. \citealt*{Myles2021}; \citealt{kids450}). In this work we present a method for characterizing such an ensemble of PDFs that collectively represent the knowledge of the redshift distribution for a galaxy sample.

Measuring and quantifying the uncertainty of redshift distributions often involves detailed studies of simulated galaxy catalogs where particular sources of error can be tightly controlled. For example, simulation codes easily facilitate changes in the number and spatial extent of galaxies used, biases in the assumed distribution of true galaxy redshifts, and the level of photometric noise in the survey. In this work, we present a methodology for mapping the variation present in an ensemble of redshift distributions measured in simulations to redshift distributions measured from the data, and vice versa. Our methodology relies on probability integral transformations to transfer the variation in an ensemble of distributions to another fiducial distribution. We call this method PITPZ for the probability integral transformations (PITs) that characterize and enable it and for the redshift `$z$' distributions that it is designed to help estimate. Although this method is designed and discussed in the context of relating effects measured in cosmological simulations to analogous measurements on data, its potential for application is notably broader than this.

This paper is organized as follows: in \S \ref{sec:method} we describe the PITPZ method and its differences compared to related existing methods, in \S \ref{sec:implementation} we discuss how we implement our method as software, in \S \ref{sec:conservation} we derive quantities conserved by the transformations of the method, in \S \ref{sec:simulated_likelihood} we show an example use of this method for propagating photometric calibration uncertainty to redshift distributions of galaxies in the Dark Energy Survey, in \S \ref{sec:results} we show results of the experiment outlined in \S \ref{sec:simulated_likelihood}, and in \S \ref{sec:conclusion} we conclude.

A flat $\Lambda$CDM cosmology with $H_0= 70 $ km s$^{-1}$ Mpc$^{-1}$ and $\Omega_{\text{m}} = 0.3$ is assumed throughout this work. Other cosmological parameters are taken to be consistent with \textit{Planck} 2018 $\Lambda$CDM cosmology \cite{planck2018}.
\section{Method} 
\label{sec:method}
\begin{figure*}
\centering
\includegraphics[width=\linewidth]{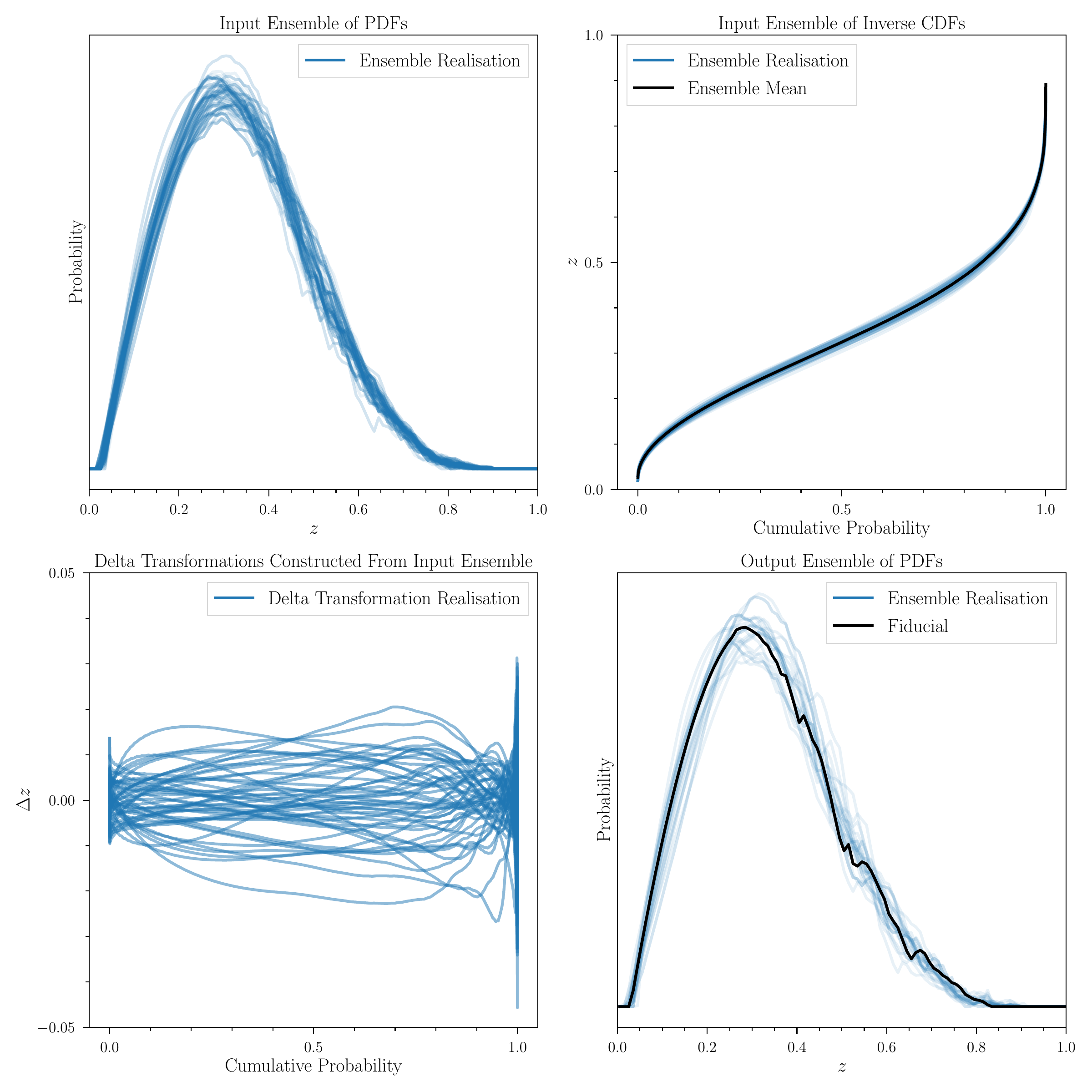}
\caption{PITPZ method used to propagate the uncertainty associated with the mock ensemble shown in the top left panel onto the mock fiducial curve of the bottom right panel. \textit{Top left}: Input ensemble of PDFs. The variation between these curves is the information we want to transfer. \textit{Top right}: Input ensemble of CDFs. \textit{Bottom left}: Delta transformations constructed from the input ensemble by taking the difference of inverse CDFs with respect to the mean inverse CDF. \textit{Bottom right}: Output ensemble of PDFs constructed by applying delta transformations to the inverse CDF of the fiducial \pz, then converting the result to a PDF.} \label{fig:pit_construction_and_method}
\end{figure*}

This section describes the PITPZ method for transferring the variation measured in one ensemble of distributions to another distribution. We provide a visual illustration of the method in Figure \ref{fig:pit_construction_and_method} to accompany the text of this section. 

In our description of the PITPZ method we use notation \pz to denote the probability distribution function of a random variable $z$ of interest. In this work, the variable of interest is galaxy redshift for a weak lensing sample of galaxies, but we refer only to abstract general probability distributions in \S \ref{sec:method}, \ref{sec:implementation}, and \ref{sec:conservation} because our method is broadly applicable to any problem with an ensemble of probability distribution functions describing some uncertainty. We thus defer specific redshift discussion until the analyses discussed in the sections thereafter.

PITPZ requires two inputs and produces one output. Namely, the two inputs are:

\begin{enumerate}[label=\Roman*., labelwidth=*]
\item A fiducial \pz measurement or ensemble of measurements. We denote this ensemble with \pzfid. While only one such measurement is needed for the purposes of this algorithm, the algorithm accommodates having an ensemble of fiducial \pz measurements to, for example, sequentially propagate multiple independent sources of uncertainty.
\item An ensemble of redshift distributions whose variation we want to map to \pzfid. We call this ensemble the input ensemble and denote it with \pzAi, where $i$ is an index for each realisation in the ensemble. 
\end{enumerate}

The sole output is:
\begin{enumerate}[resume, label=\Roman*.]
    \item An ensemble of \pz whose variation is related to the variation between realisations of the input ensemble but which is mapped onto \pzfid. We call this ensemble the output ensemble and denote it with \pzB. We describe quantitatively the relationship between the variation of the input ensemble and the variation of the output ensemble in \S \ref{sec:conservation}.
\end{enumerate}

We begin by computing the inverse cumulative distribution function (inverse CDF, also called the quantile function) $F^{-1}_i$ for each realisation \pzi in the input ensemble. This can be written as
\begin{equation}
    F^{-1}_{i}(p) = \{z :  F_{i}(z) = p \} \quad 
\end{equation}

where the CDF is defined as 

\begin{equation}
    \quad F_{i}(z) = \int_{-\infty}^{\infty} p_{i}(z^{\prime}) dz^{\prime} = \int_{0}^{z_{\text{max.}}} p_{i}(z^{\prime}) dz^{\prime} \; .
\end{equation}
The integral transforming \pz to $F(z)$ is called a probability integral transformation \citep{dodge2006oxford}. Our method relies on these transformations to generate the cumulative distribution functions necessary to subsequently produce a transformation that transfers variation from the input ensemble onto \pzfid. 

We note that our method, while making use of PITs, differs from past uses of PITs for galaxy redshift estimation. Such past work includes the use of PITs to assess redshift biases by taking advantage of the fact that the PIT of a proper PDF is uniformly distributed, so deviations from uniform distributions in PITs computed from redshift PDFs indicate the presence of biases in these underlying PDFs (see e.g. \citealt{Bordoloi2010, Polsterer2016, Freeman2017, Tanaka2018, Schmidt2020, Shuntov2020, Hasan2022, Zhang2022}). Our method, by contrast, uses PITs to construct another transformation entirely which is used to alter \pz to make them more like some other \pz, as to be described in greater detail in the following text.

We define a new transformation which we call a delta transformation (denoted here as $T$) as the difference between the inverse CDF $F^{-1}_{i}$ of a given realisation  in the input ensemble and the \textit{average} inverse CDF of the input ensemble:
\begin{equation}
    \label{eqn:pit}
    T_{i} = F^{-1, \mathrm{in.}}_{i} - \langle F^{-1, \mathrm{in.}} \rangle.
\end{equation}

Given this definition, each delta transformation encodes the difference between a given realisation of the input ensemble and the mean of the realisations of said input ensemble.
We apply these transformations by adding each delta transformation to the inverse CDF $F^{-1}_{\text{fid.}}$ of the fiducial data $n(z)$:
\begin{equation}
    \label{eqn:pit_apply}
    F^{-1, \mathrm{out.}}_{i} = F^{-1, \text{fid.}}_{i} + T_{i}.
\end{equation}
Given this ensemble of transformed inverse CDFs of \pzfid, we construct the output ensemble by taking the inverse of these inverse CDFs to yield CDFs, then differentiating to yield PDFs:
\begin{equation}
\label{eqn:pit_differentiate}
    p_{i}^\mathrm{out.}(z) = \dfrac{d}{dz} \left(F_{i}^\mathrm{out.}\right).
\end{equation}
\section{Implementation} 
\label{sec:implementation}
The conceptual algorithm described in \S \ref{sec:method} for generating an ensemble of \pz involves manipulating smooth probability density and quantile functions. We circumvent implementation complications that arise from operating on smooth functions by \textit{evenly sampling} each PDF to generate an ordered list of \textit{n} samples $\{z_0 ... z_n\}$ from each \pz and manipulating these samples, rather than the quantile functions directly as follows. In practice the probability density functions used are often stored digitally as histograms, in which case our sampling procedure avoids complications related to differing normalizations and bin size and range.

In brief, applying a delta transformation (as in Equation \ref{eqn:pit_apply}) amounts to generating an ordered list of samples ${z}$ from each \pz, adjusting the values of those samples with the delta transformation, and computing the distribution of the adjusted samples for a specified histogram binning. We first determine the number of samples to be apportioned to each histogram bin, then use those samples to compute and apply each delta transformation, and finally compute the new \pz from each array of ordered, adjusted samples.

We use the largest remainder method to apportion the discrete samples among histogram bins as closely to the bins' relative probability as is possible \citep{Tannenbaum2010}. This method consists of dividing the total number $n$ of samples to be apportioned by the histogram value \pz of each histogram bin. Each bin is apportioned a number of samples equal to the integral part of its respective quotient. The histogram bins are then ranked by the size of their remainders, and each bin is assigned an additional sample until the remaining samples have been fully allocated. This procedure is done for the fiducial distribution \pz and for each realisation \pzAi constituting the input ensemble. After using this method to compute the appropriate number of samples apportioned to each bin, we distribute those samples evenly across the width of the bin. This yields the following sets of ordered redshift values:
\begin{enumerate}[label=\Roman*., labelwidth=*]
    \item 1 (or more) set $\{z_0, z_1, ..., z_{n}\}^{\text{fid.}}$
    \item N sets $\{z_0, z_1, ..., z_{n}\}^{\mathrm{in.}}_{i}$
\end{enumerate}
Here the $j$th value $z_{i,j}$ of the $i$th set of ordered redshift samples $\{z\}_i$ represents the redshift corresponding to the $\frac{j}{n}$th quantile of the distribution. In other words, these samples constitute the quantile function for \pz. 

We then compute the delta transformations by taking the difference of each ordered sample of a realisation in the input ensemble and the corresponding ordered sample for the mean of these realisations:
\begin{equation}
    \label{eqn:pit_unnormalized_implementation}
    \begin{split}
    &T_{i} = \\ 
    &= \{z_0, z_1, ... ,z_{n}\}_{i}^{\mathrm{in.}} - \{z_0, z_1, ... z_{n}\}^{\langle \mathrm{in.} \rangle}\\
    &= \{\Delta_0, \Delta_1, ... \Delta_{n}\}_i
    \end{split}
\end{equation}
Applying these delta transformations amounts to adding each of these $\Delta z$ values to the value of its corresponding quantile in the list of ordered samples of \pzfid. For a single delta transformation $T = \{ \Delta_0, \Delta_1, ... , \Delta_{n}\}$, the implementation of Equation \ref{eqn:pit_apply} is then:
\begin{equation}
    \label{eqn:pit_apply_implementation}
    \begin{split}
    \{ z_0^\mathrm{out.}, z_1^{\mathrm{out.}}, ... z_n^{\mathrm{out.}} \} = \{ z_0^{\mathrm{fid.}} + \Delta_0, z_1^{\mathrm{fid.}} + \Delta_1, ... z_n^{\mathrm{fid.}} + \Delta_n\}
    \end{split}
\end{equation}

We note that as a result of the delta transformation some samples can be shifted outside of the range of acceptable values, e.g. below zero in the case of cosmological redshift. In the case of redshift distributions we discard these samples and increase the value of the remaining samples such that the mean redshift of the distribution is not changed. Once we have the perturbed samples described by Equation \ref{eqn:pit_apply_implementation}, constructing the final modified \pz is done by binning the samples with any given histogram bin edges, which is done in lieu of Equation \ref{eqn:pit_differentiate}.
\section{Conservation Rules of Delta Transformations}
\label{sec:conservation}
Recall the goal of the PITPZ method: we aim to propagate uncertainties to measured redshift distributions. Past analyses have used coherent shifts of measured redshift distributions to lower and higher values, with the shifts drawn from a Gaussian distribution whose standard deviation encapsulates mean redshift uncertainty (see e.g. \citealt{Hoyle2018}). This approach produces an output ensemble of PDFs that only varies in mean redshift, but in reality many sources of uncertainty produce more complicated variations than simple mean shifts. The goal of PITPZ is to preserve the full correlation structure across an input ensemble in a constructed output ensemble. This section is dedicated to illustrating how this information is conserved by the PITPZ method.

\begin{figure*}
\centering
\includegraphics[width=\linewidth]{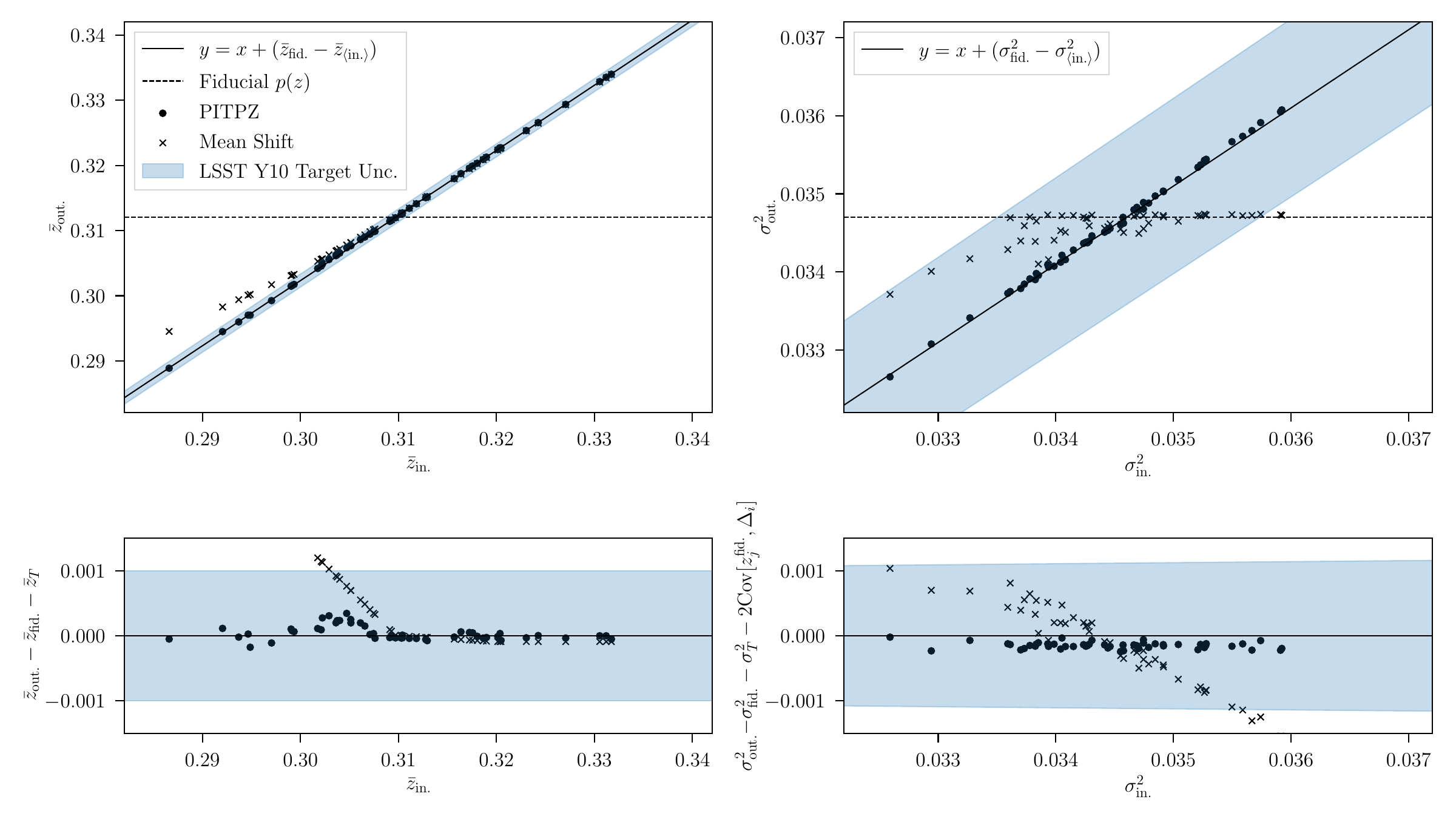}
\caption{Illustration of mean and variance conservation by the PITPZ method and of improved behavior compared to mean shifts. Shown here are results for the first tomographic bin of the experiment described in \S \ref{sec:simulated_likelihood}. \textit{Top}: Relationship in redshift distribution moments between the input ensemble and output ensemble realisations. 
\textit{Bottom}: Deviations from the conservation rules derived in \S \ref{sec:conservation} due to numerical noise in our software implementation of the formalism described. The blue uncertainty region corresponds to the LSST Y10 WL analysis uncertainty requirements of $0.001(1+z)$ on the mean and $0.003(1+z)$ on the standard deviation (here scaled to variance) of redshift at $z=0$ \citep{lsst_desc_sci_req}.} \label{fig:pit_info_conservation}
\end{figure*}

Recall that the starting point for applying the PITPZ method is two inputs: a fiducial measured \pzfid (or an ensemble of such fiducial measurements) and an input ensemble \pzAi of redshift distributions whose variation encodes uncertainty due to some relevant effect(s). Our algorithm produces an output ensemble \pzBi which has mapped the variation in the input ensemble onto the fiducial measurement \pzfid. Posing the question of information conservation in the broadest possible sense, we want to relate each central moment of each realisation in \pzAi to the corresponding central moment of its counterpart realisation in \pzBi. We proceed by deriving the conservation rules for the mean, variance, and skewness of a realisation of the output ensemble in terms of the corresponding moments of the fiducial \pz, the realisation of the input ensemble used, and the mean of the realisations of the input ensemble. Figure \ref{fig:pit_info_conservation} shows the performance of our software implementation of PITPZ to conserve the rules derived for mean and variance. Inspection of this figure illustrates that PITPZ produces an output \pz realisation whose mean differs from the fiducial in proportion to how the mean of the corresponding realisation of the input ensemble differs from the mean of the input ensemble. By contrast, mean shifts maintain this relationship only when sufficiently far from the edges of the allowed parameter limits. The fact that the observed numerical noise lies within the LSST uncertainty region illustrates that the deviation from conservation of the mean value is negligible for near-term weak lensing redshift calibration applications. PITPZ preserves a similar relationship for the variance, but mean shifts do not transfer the relative change in width of realisations in the input ensemble to the constructed output ensemble. Although for the source of uncertainty propagated for this figure (see \S \ref{sec:simulated_likelihood}) the changes in \pz width introduced by the mean shift method are within the LSST Year 10 target uncertainty, it is the combined value for all sources of uncertainty that should be ultimately compared to the target error budget. In practice, using PITPZ may be necessary to meet the LSST Year 10 target uncertainties.

In this section we introduce the following notation convention: Overlines represent averages over the redshift value samples, which are indexed with $j$. For example, the mean redshift $z$ of \pz is represented by $\overline{z}$. Brackets represent averages over the redshift distribution realisations of an ensemble, which are indexed by $i$. For example, the mean \pz of the input ensemble, \pzAi, is represented by \pzAavg.

\subsection{Mean of Redshift Distributions}
Measuring the mean redshift of each constituent realisation of the input ensemble yields a distribution of mean redshifts $\bar{z}_{i}^\mathrm{in.}$ where $\bar{z}_{i}^\mathrm{in.} = \int z' p_{i}^{\mathrm{in.}}(z')dz'$. We aim to derive the relation between each mean redshift in this ensemble and the mean redshift of the corresponding output in the output ensemble produced by the PITPZ algorithm, $\bar{z}_{i}^\mathrm{out.}$.

As introduced in \S \ref{sec:implementation}, we can represent a given realisation of the input ensemble \pzA, a given delta transformation $T$, and the resulting realisation of the output ensemble \pzB as a set of ordered samples:

\begin{equation}
\begin{split}
    p^{\mathrm{in.}}(z) &\Leftrightarrow \{ z_0^\mathrm{in.}, z_1^\mathrm{in.}, ... z_n^\mathrm{in.} \}\\
    T &\Leftrightarrow \{ \Delta_0, \Delta_1, ... \Delta_n \}\\
    p^{\mathrm{out.}}(z) &\Leftrightarrow \{ z_0^{\mathrm{out.}}, z_1^{\mathrm{out.}}, ... z_n^{\mathrm{out.}} \} \\
    &\ \ \ \ = \{ z_0^{\mathrm{fid.}} + \Delta_0, z_1^{\mathrm{fid.}} + \Delta_1, ... z_n^{\mathrm{fid.}} + \Delta_n\}
    \end{split}
\end{equation}

It is straightforward to prove that the mean redshift of each realisation of the output ensemble is the sum of the mean redshift of the fiducial \pz and the mean value of the shifts comprising the delta transformation. In the following we use our customary labels of `$\mathrm{in.}$' and `$\mathrm{out.}$' to represent single realisations of the input and output ensembles, respectively, and the letter $T$ to likewise represent a single delta transformation. With this convention, each input-output pair follows the following conservation rule:

\begin{equation}
\begin{split}
    \bar{z}^{\mathrm{out.}} &= \frac{1}{n} \sum_j^n z^{\mathrm{out.}}_j \\
    &= \frac{1}{n} \sum_j^n \left( z^{\mathrm{fid.}}_j + \Delta_{j} \right) \\
    &= \frac{1}{n} \sum_j^n z^{\mathrm{fid.}}_j + \frac{1}{n} \sum_j^n \Delta_{j} \\
    &= \bar{z}^\mathrm{fid.} + \bar{\Delta} \\
    &= \bar{z}^\mathrm{fid.} + \bar{z}^\mathrm{\mathrm{in.}} - \bar{z}^{\langle \mathrm{in.} \rangle}
\end{split}
\end{equation}

\subsection{Higher order moments of Redshift Distributions}
We present results for the variance and skewness here, deferring the full derivation to Appendix \ref{sec:conservation_appendix}.

Our expression for the variance of a realisation in the output ensemble is

\begin{equation}
\label{eqn:std_b}
\begin{split}
    \sigma^2_{\mathrm{out.}} &= \sigma_{\mathrm{fid.}}^2 + \sigma_T^2 + 2 \ \mathrm{Cov}[z_j^\mathrm{fid.}, \Delta_j]\\
    &= \sigma_{\mathrm{fid.}}^2 + \sigma_{\mathrm{in.}}^2 + \sigma_{\langle \mathrm{in.} \rangle}^2 - 2\  \mathrm{Cov} \left[ z_j^{\mathrm{in.}}, z_j^{\langle \mathrm{in.} \rangle} \right] + 2 \ \mathrm{Cov}[z_j^\mathrm{fid.}, \Delta_j]
\end{split} 
\end{equation}

Our expression for the skewness of a realisation in the output ensemble is:

\begin{equation}
\begin{split}
    \sigma_{\mathrm{out.}}^3 \widetilde{\mu}_{3}^{\mathrm{out.}} & = \sigma_{\mathrm{fid.}}^3 \widetilde{\mu}_{3}^{\mathrm{fid.}} + \sigma_{\mathrm{in.}}^3 \widetilde{\mu}_{3}^{\mathrm{in.}} + \sigma_{\langle \mathrm{in.} \rangle}^3 \widetilde{\mu}_{3}^{\langle \mathrm{in.} \rangle}\\
    &+ 3 \sigma_{\mathrm{fid.}}^2 \sigma_{\mathrm{in.}} S(z_j^{\mathrm{fid.}}, z_j^{\mathrm{fid.}}, z_j^{\mathrm{in.}}) \\
    &- 3 \sigma_{\mathrm{fid.}}^2 \sigma_{\langle \mathrm{in.} \rangle} S(z_j^{\mathrm{fid.}}, z_j^{\mathrm{fid.}}, z_j^{\langle \mathrm{in.} \rangle})\\
    &+ 3 \sigma_{\mathrm{fid.}} \sigma_{\mathrm{in.}}^2 S(z_j^{\mathrm{fid.}}, z_j^{\mathrm{in.}}, z_j^{\mathrm{in.}}) \\
    &- 6 \sigma_{\mathrm{fid.}} \sigma_{\mathrm{in.}} \sigma_{\langle {\mathrm{in.}} \rangle} S(z_j^{\mathrm{fid.}}, z_j^{\mathrm{in.}}, z_j^{\langle \mathrm{in.} \rangle})\\
    &+ 3 \sigma_{\mathrm{fid.}} \sigma_{\langle \mathrm{in.} \rangle}^2 S(z_j^{\mathrm{fid.}}, z_j^{\langle \mathrm{in.} \rangle}, z_j^{\langle \mathrm{in.} \rangle}) \\
    &- 3 \sigma_{\mathrm{in.}}^2 \sigma_{\langle \mathrm{in.} \rangle} S(z_j^{\mathrm{in.}}, z_j^{\mathrm{in.}}, z_j^{\langle \mathrm{in.} \rangle})\\
    &+ 3 \sigma_\mathrm{in.} \sigma_{\langle \mathrm{in.} \rangle}^2 S(z_j^{\mathrm{in.}}, z_j^{\langle \mathrm{in.} \rangle}, z_j^{\langle \mathrm{in.} \rangle})
    \end{split}
\end{equation}

where the $S$ denotes the coskewness of three random variables $X$, $Y$, and $Z$:
\begin{equation}
S(X,Y,Z) = \frac{\text{E}[(X - \text{E}(X)) (Y - \text{E}(Y)) (Z - \text{E}(Z))]}{\sigma_X \sigma_Y \sigma_Z}
\end{equation}

\section{Cosmological Impact Analysis}
\label{sec:simulated_likelihood}
Having defined PITPZ as a statistical method and illustrated the rules by which it conserves and transfers information from one distribution of PDFs to another, we now turn to understanding how this can affect scientific conclusions in the context of weak lensing cosmology experiments. For the remainder of this work, we choose to denote our probability distribution function of interest as \nz to remain consistent with the redshift calibration literature, in which \nz represents a weighted number density of galaxies at redshift $z$ where each galaxy's may be weighted according to its contribution to the associated shear catalog (for more information about weight choices see e.g. \citealt{y3-shapecatalog}). We note that \nz has a different normalization than the probability density function of a galaxy in the survey having a specific redshift and emphasize that \nz is not the probability distribution function for the redshift of an individual galaxy. 

Weak gravitational lensing refers to the accumulated deflections to the path of light from a distant source galaxy as it travels through the large-scale structure of the Universe toward an observer. In order to interpret the coherent distortions in the shapes of large samples of observed galaxies due to this effect, we must have a constraint on the redshift of the source galaxies and the intervening distribution of lensing matter. In this context, the salient question is how using PITPZ to generate \nz realisations whose variation encodes uncertainties in the redshift distributions of the selected galaxy sample will affect the uncertainty on parameters of the cosmological model being tested with weak lensing analyses. In practice, the relationship between variations of \nz realisations and cosmology uncertainty is that evaluating the cosmology likelihood function given weak lensing data should sample over an ensemble of \nz realisations. For the purpose of our work, the question of how \nz uncertainty and cosmology are related can be reduced to assessing the impact that using PITPZ to construct redshift distributions has on the resulting distribution of lensing signal amplitude (for a given lens redshift). To this end we first briefly summarize the way galaxy photometry is used in the redshift calibration scheme applied in this work, deferring to \citet*{Myles2021} for a full description.

\subsection{DES Year 3 Redshift Methodology}
The DES Y3 redshift calibration relies on a method called \sompz developed to take advantage of the DES deep-drilling fields where longer exposure times and spatial overlap with near-infrared surveys provides more information to use for redshift inference \citep*{Buchs2019, Myles2021, Hartley2022}. In this method, the deep-field galaxies serve as an intermediary between galaxies with secure (e.g. spectroscopic) redshifts and the overall wide-field sample; the deep-field galaxies play the crucial role of enabling secure redshifts to be used for subsamples of galaxies while avoiding selection bias between the secure redshift sample and galaxies in the overall wide-field survey sample (for more information on such selection bias, see \citealt{GruenBrimioulle2017}). Within this scheme, redshift distributions are computed in small regions of deep-field color-magnitude space. The wide-field galaxy density is determined in small regions of wide-field color-magnitude space. The ultimate calibrated redshift distributions of the wide-field sample are the weighted sum of redshift distributions in deep-field color-magnitude space, where weights are the likelihood of given deep galaxies being detected and selected in the wide-field sample as determined using the \balrog image simulation package \citep{Everett2020}. \sompz is additionally combined with independent information from galaxy clustering and shear ratios \citep*{Myles2021, y3-sourcewz, y3-shearratio}. The final product of this kind of redshift calibration is not a single \nz, but rather an ensemble of \nz whose variations encode the uncertainty. This ensemble can be used in cosmology analyses by sampling the ensemble for each evaluation of the cosmological likelihood function. PITPZ is designed as a method for generating such an ensemble to be sampled in cosmology analyses.

\subsection{Experimental Design}

Among the several sources of uncertainty inherent to the DES Year 3 redshift methodology, the photometric calibration of the deep-field galaxies stands out due to the novel use of these galaxies to improve our calibration. This uncertainty is best understood by taking advantage of realistic simulations in which photometric calibration error can be easily scaled at will. We therefore choose this source of uncertainty to illustrate the characteristics of our PITPZ method for propagating uncertainty.

\begin{figure*}
\centering
\makebox[\textwidth][c]{\includegraphics[width=1.3\linewidth]{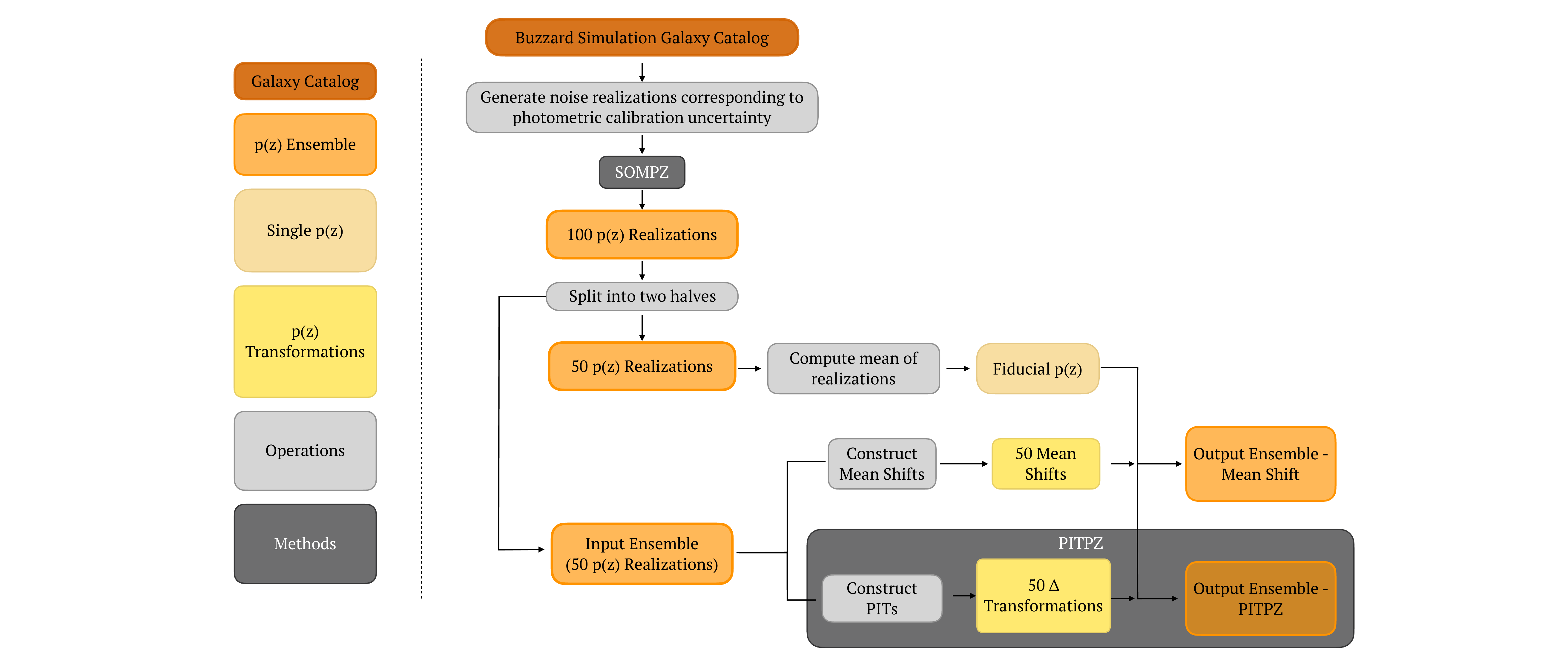}}
\caption{Illustration of the experimental design of the cosmological impact analysis in this work. The input ensemble is produced by running \sompz 50 times with varying deep-field photometric zero-points. The fiducial \nz is produced by taking the mean of an ensemble produced by running \sompz 50 times again with varying deep-field photometric zero-points. `Output Ensemble -- Mean Shift' is constructed by shifting the fiducial \nz by the mean value of each PIT; `Output Ensemble -- PITPZ' is constructed with the PITPZ method, i.e. by applying the full-shape delta transformations constructed from the input Ensemble to alter the fiducial \nz.}  \label{fig:sim_like_experimental_design}
\end{figure*}

Our experimental design to illustrate the impact of PITPZ consists of the procedure described in the following test and illustrated in Figure \ref{fig:sim_like_experimental_design}. 

We begin with an ensemble of 100 \nz produced using the \buzzard simulations \citep{Derose2019} where each realisation has zero-point offsets according to the photometric calibration uncertainty measured by \citet*{Hartley2022} are introduced to the deep-field photometry. The variation between the \nz realisations in this ensemble reflects the uncertainty in \nz due to deep-field photometric zero-point uncertainty. 

We split this ensemble into two halves of 50 realisations each. The first half is used to construct delta transformations relative to the mean. Because it is used in this way, the first half serves the role of the input ensemble as defined in \S \ref{sec:method}, so it is labelled \nzA. The second half is to construct the fiducial \nzfid: \nzfid is simply the mean of the \nz comprising the second half. 

We apply the delta transformations made from the first half (i.e. from the input ensemble) to this fiducial \nzfid. As an alternative to applying the delta transformations, we also apply to the fiducial \nzfid the mean shifts corresponding to the difference in mean redshift between each realisation of the input ensemble and the mean of the realisations of the aforementioned input ensemble; this is a simpler alternative to PITPZ which has been employed for past redshift calibration analyses, e.g. \citealt{Jee2013, Bonnett2016}; \citealt*{Hoyle2018}. As a result, we have produced two versions of the output ensemble: one with PITPZ and one with mean shifts. The mean shift ensemble transfers only changes in the mean redshift between realisations in \nzA; by contrast PITPZ transfers the information for higher than mean-order moments according to the conservation rules shown in \ref{sec:conservation}. In short, PITPZ transfers the full correlation structure of the realisations generated by the simulations. These two versions of the output ensemble should have transferred a different aspect or `amount' of information from \nzA to \nzfid. The difference between these two versions of the output ensemble will demonstrate the benefits of using PITPZ rather than mean shifts. To summarize, the three \nz ensembles discussed are: 

\begin{enumerate}[label=\Roman*.]
    \item (Input Ensemble):  First determine random zero-point offsets due to the uncertainty of the photometric calibration error by drawing from a Gaussian centred on zero with standard deviation set to the uncertainty of the deep field photometric calibration in each band. Shift all deep field magnitudes according to the result of this draw in each respective band for each deep field. Use these altered deep-field magnitudes as input to a run of the \sompz method on the \buzzard simulated galaxy catalogs. Select the first 50 realisations and construct delta transformations from them.
    
    \item (Output Ensemble -- Mean Shift): \nz constructed by applying mean shifts (rather than full-shape delta transformations) to the fiducial \nz. 
    
    \item (Output Ensemble -- PITPZ): \nz constructed by applying full-shape delta transformations to the fiducial \nz. Following the notation of \S \ref{sec:method}, this ensemble is labelled \nzB.
\end{enumerate}

\begin{figure*}
\centering
\includegraphics[width=\linewidth]{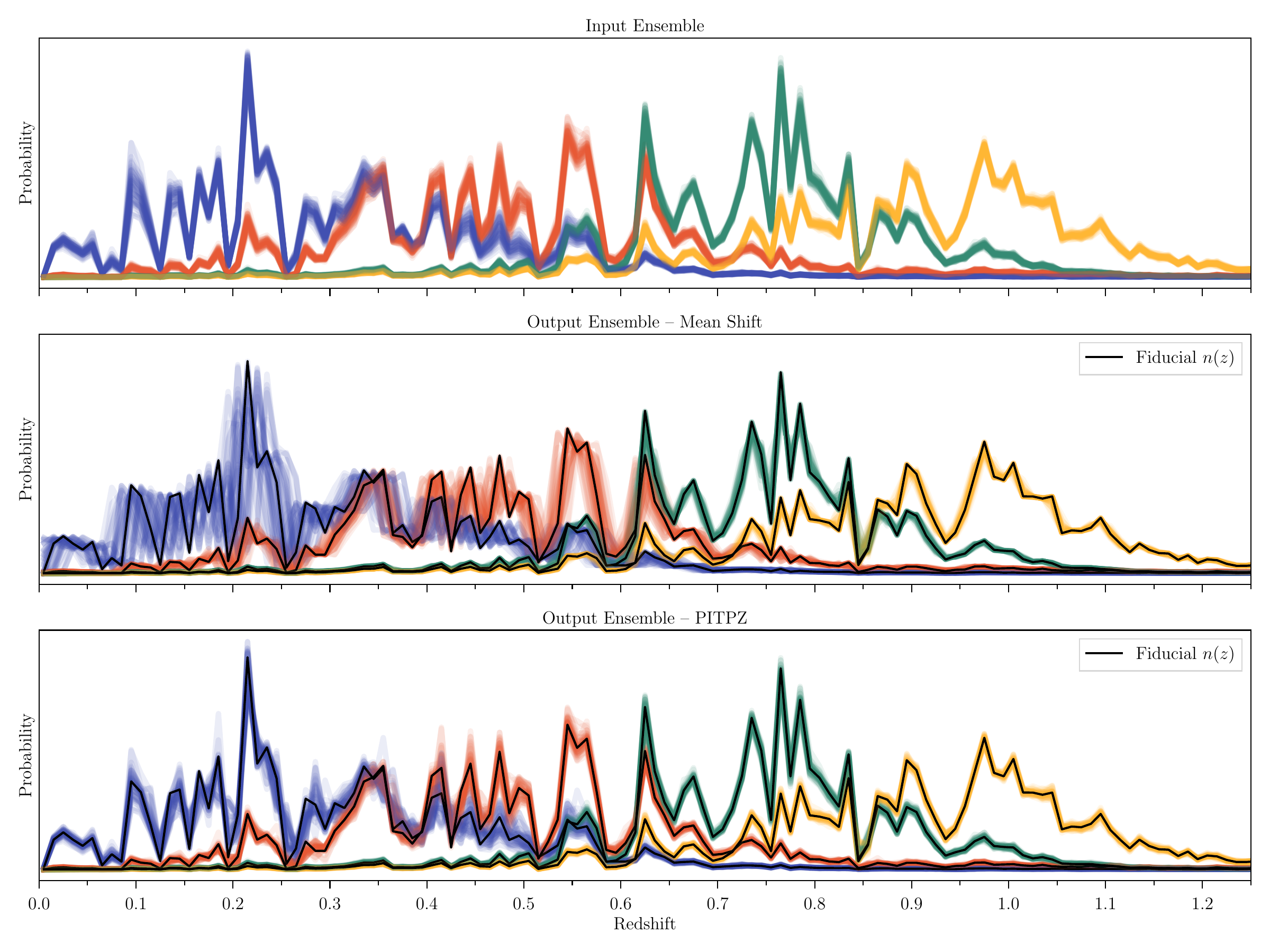}
\caption{Illustration of the \nz distributions used in the simulated likelihood analysis in this work. The input ensemble is produced by running \sompz 50 times with varying deep-field photometric zero-points. `Output Ensemble -- Mean Shift' is constructed by shifting the fiducial \nz by the mean value of each PIT; `Output Ensemble -- PITPZ' is constructed with the PITPZ method, i.e. by applying the full-shape Delta Transformations constructed from the input ensemble to alter the fiducial \nz.}  \label{fig:nz_ensembles_all}
\end{figure*}

These \nz are shown in Figure \ref{fig:nz_ensembles_all}. With these mock redshift distribution ensembles produced, we turn to assessing the difference between them for cosmology analysis. Our analysis consists in computing the uncertainty on the lensing amplitude associated with each ensemble, which relates closely to uncertainty on cosmological parameters.   

We are interested in the following comparisons of the lensing amplitude distribution results yielded from these analyses:

\begin{enumerate}[label=\arabic*.]
    \item The difference between the lensing amplitude distributions associated with II and III illustrates the residual effect on redshift distributions of zero-point uncertainties beyond the first-order shift of the mean redshift. This is equivalent to illustrating the importance of using PITPZ, rather than simpler mean shifts, to incorporate this systematic uncertainty into redshift distributions.
    \item Because the input ensemble serves as a ground truth for the degree of variation due to photometric calibration uncertainty present in the simulations, any difference between the lensing amplitude distributions associated with I and III illustrates the residual effect on redshift distribution of zero-point uncertainties beyond what is corrected for with delta transformations produced with \buzzard. This is equivalent to illustrating the impact of higher than first-order moments due to the effect of photometric calibration uncertainty beyond what can be accounted for with the PITPZ method. In summary, any difference here illustrates shortcomings of the PITPZ method. 
\end{enumerate}

While the primary goal of this work is the illustration of the importance of using the delta transformation to preserve higher-order information than lower $n$-th order statistics in generating ensembles of probability distributions (i.e. comparison 1), this experimental design facilitates a secondary goal of illustrating the impact of our chosen source of uncertainty -- photometric calibration error -- on cosmology constraints. This secondary goal can play a role in informing future observing strategy decisions to collect the data necessary to reduce this uncertainty.

It remains to describe the relevant statistic that relates redshift distributions to constraints on the parameters of a given cosmological model. In practice, weak gravitational lensing involves inferring the matter distribution from coherent distortions in the measured shapes of galaxies. The presence of tangential alignment in galaxy shapes measured on the sky corresponds to the presence of a matter overdensity along the line of sight. The observed mean tangential shear $\gamma_t$ associated with a separation angle $\theta$ on the sky can be expressed in terms of the lensing \textit{convergence} that describes the amount of lensing

\begin{equation}
    \langle \gamma_t \rangle (\theta) = \overline{\kappa} (< \theta) - \langle \kappa \rangle (\theta).
\end{equation}

Convergence, in turn, can be written in terms of the total projected mass density $\Sigma$ along a line-of-sight $\vec{\theta}$ and a \textit{critical surface density} parameter which characterizes the lensing system

\begin{equation}
    \kappa(\vec{\theta}) \equiv \frac{\Sigma (\vec{\theta})}{\Sigma_{\mathrm{crit.}}}.
\end{equation}

This critical surface density due to lensing of a source at distance $D_{\mathrm{s}}$ from the observer by a lens (i.e. deflector) at distance $D_{\mathrm{d}}$ from the observer, in a universe where the distance between the source and the lens is $D_{\mathrm{ds}}$, is defined as follows under the assumption that the distances between source, lens, and observer are all much greater than the spatial extent of the lens (see e.g. \citealt{BartlemannSchneider2001})

\begin{equation}
    \sigmacritinvmm \equiv \frac{c^2}{4\pi G} \frac{D_\mathrm{s}}{D_{\mathrm{d}} D_{\mathrm{ds}}}.
\end{equation}

This definition illustrates that uncertainty on galaxy distance corresponds directly to uncertainty on critical surface density, which in turn directly limits the degree to which projected mass density and therefore cosmology can be constrained. For this reason we choose critical surface density to test the impact of PITPZ on cosmology. 

The shear $\gamma(\vec{\theta}, z_s) $ to which a particular source galaxy image is subject is a function of source galaxy redshift, so the mean shear observed along a line of sight $\vec{\theta}$ must be expressed with respect to the source galaxy redshift distributions \citep{MacCrann2022, y3-cosmicshear1}

\begin{equation}
    \gamma(\mathbf{\vec{\theta}}) = \int dz_{\mathrm{s}} n(z_{\mathrm{s}}) \gamma(\mathbf{\vec{\theta}}, z_{\mathrm{s}}).
\end{equation}

Similarly, the total averaged lensing signal amplitude can be expressed in terms of the critical surface density integrated in the same way as the total shear

\begin{equation}
\label{eqn:sigmacritinv}
    \avgsigmacritinvmm = \frac{\int_{z_{\mathrm{l}}}^{z_{\mathrm{s}, \mathrm{max.}}}{\frac{4 \pi G}{c^2} \frac{ D_{\mathrm{d}}(z_\mathrm{l}) D_{\mathrm{ds}} (z_{\mathrm{s}}, z_{\mathrm{l}})}{D_{\mathrm{s}} (z_{\mathrm{s}})} n(z_{\mathrm{s}}) dz_{\mathrm{s}}} }{\int_{0}^{z_{\mathrm{s}, \mathrm{max.}}} n(z_{\mathrm{s}}) dz_{\mathrm{s}}}
\end{equation}

where the denominator is a normalization factor. Here $D_\mathrm{d}$, $D_\mathrm{s}$, and $D_{\mathrm{ds}}$ are determined by the lens and source redshifts $z_{\mathrm{l}}$ and $z_{\mathrm{s}}$. Equation \ref{eqn:sigmacritinv} is a statistic to relate uncertainty on \nz to uncertainty on cosmology results. Note that this statistic is a weighted integral of \nz, and effectively measures the probability density at redshift higher than the lens redshift $z_\mathrm{l}$, with higher redshift probability being weighted higher. As such, this statistic depends on higher than mean-order moments in \nz. While mean redshift is the most important determining factor in the value of this statistic, at fixed mean redshift increasing variance, for example, will increase the probability at the highest redshifts. As a result, we expect this quantity to be more accurately evaluated from \nz constructed with PITPZ than from simpler mean shifts because PITPZ propagates uncertainty to higher-order moments (c.f. Fig. \ref{fig:pit_info_conservation}). 

We compute the distribution in \avgsigmacritinv for each of our redshift distribution ensembles using the \texttt{lenstronomy} \citep{lenstronomyI, lenstronomyII} software and report the resulting values in Table \ref{tab:sim_like_analysis}. Since the uncertainty on constraints on cosmology from a cosmic shear analysis such as that conducted with the Dark Energy Survey Year 3 dataset \citep{y3-cosmicshear1, y3-cosmicshear2} is proportional to the uncertainty on lensing amplitude, the distribution of possible lensing amplitudes functions as a proxy for the resulting uncertainty on cosmological parameters. In addition to the statistic defined in Eqn. \ref{eqn:sigmacritinv}, we compute the cosmic shear two-point correlation function $\xi_{+/-}$ with each $n(z)$ in our input and output ensembles using the \texttt{CCL} package of \citet{pyccl} (for details on cosmic shear, see e.g., \citealt{y3-cosmicshear1, y3-cosmicshear2}). We integrate over this cosmic shear data vector $\xi$ and show results relating input and output values of this quantity in Fig. \ref{fig:pit_info_conservation_cosmic_shear}.

\begin{figure*}
\centering
\includegraphics[width=\linewidth]{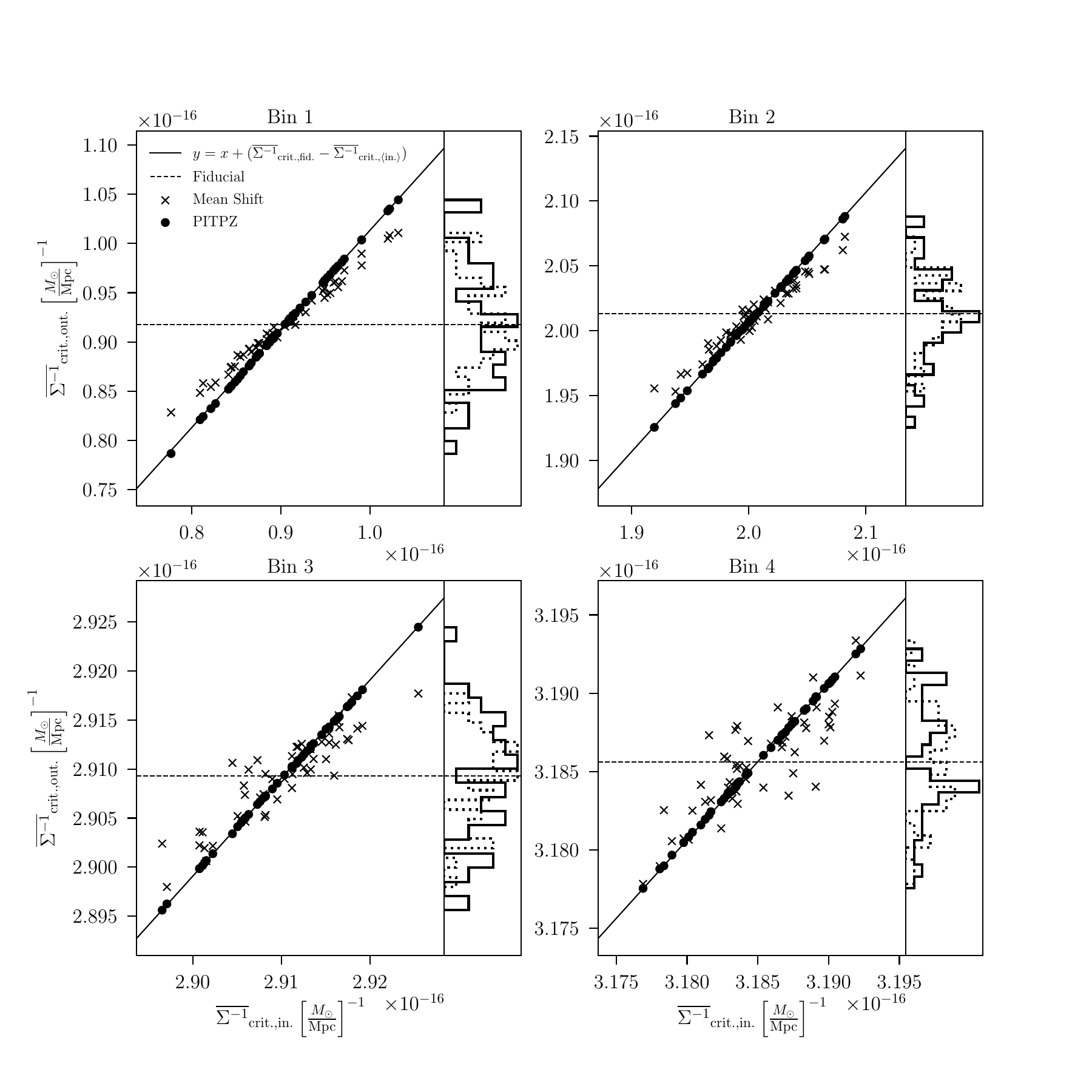}
\caption{Relationship between lensing signal amplitude in the input ensemble and the output ensemble realisations using PITPZ or mean shifts for the experiment described in \S \ref{sec:simulated_likelihood} with $z_{\mathrm{lens}} = 0.25$. We find that PITPZ more reliably transfers lensing amplitude information than mean shifts. This is explained by the fact that the lensing amplitude is a weighted integral of \nz, so higher-order moments of \nz which are conserved by PITPZ but not conserved by mean shifts will cause the mean shift to underestimate the scatter in lensing amplitude. Histograms on the side panels illustrate the distribution of lensing signal amplitude for the output ensemble, where the solid line corresponds to the output ensemble produced with PITPZ and the dotted line corresponds to that produced with mean shifts.}  \label{fig:lensing_amp_conservation}
\end{figure*}

\begin{figure*}
\centering
\includegraphics[width=\linewidth]{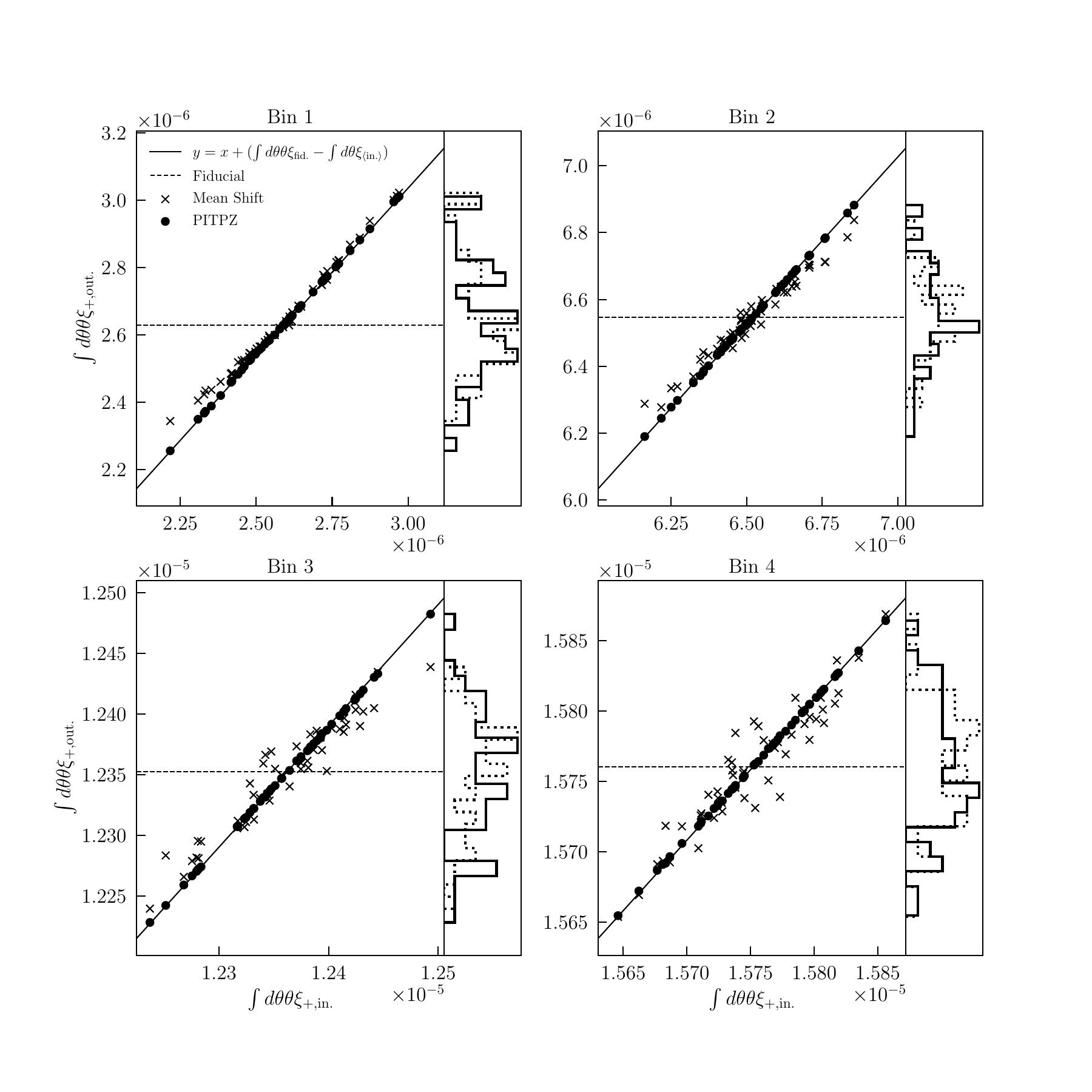}
\caption{Relationship between the cosmic shear signal amplitude as inferred from input ensemble $n(z)$ realisations to the cosmic shear signal amplitude as inferred from output ensemble $n(z)$ realisations. The output ensembles are produced with PITPZ or mean shifts with the experiment described in \S \ref{sec:simulated_likelihood}. Axis values are integrals over the full cosmic shear data vector $\xi_+$. As in Fig. \ref{fig:lensing_amp_conservation}, we find that PITPZ more reliably transfers information than mean shifts. Histograms on the side panels illustrate the distribution of signal amplitude for the output ensemble, where the solid line corresponds to the output ensemble produced with PITPZ and the dotted line corresponds to that produced with mean shifts.} \label{fig:pit_info_conservation_cosmic_shear}
\end{figure*}



\begin{table*}
\centering
\begin{tabular}{ l c c c c c c c } 

    &  & \multicolumn{2}{c}{$\overline{z}$} & \multicolumn{2}{c}{$\avgsigmacritinvmm \left[ \frac{M_{\odot}}{\mathrm{Mpc}} \right]^{-1}$} & \multicolumn{2}{c}{$\avgsigmacritinvmm \left[ \frac{M_{\odot}}{\mathrm{Mpc}} \right]^{-1}$}\\ 
    
\hline
  Name & Symbol & $\mu$ & $\sigma$ & $\mu$ & $\sigma$ & $\mu / \mu_{\text{in.}}$ & $\sigma / \sigma_{\text{in.}}$\\ 
  
\hline
 \hline
  \hline
   &  & \multicolumn{6}{c}{Bin 1}\\
\hline
Input Ensemble & $p_{\mathrm{in.}} (z)$ & 0.310 & 1.044e-02 & 9.043e-17 & 5.856e-18 & 1.000 & \textbf{ 1.000 }\\
Output Ensemble -- Mean Shift & - & 0.313 & 9.478e-03 & 9.181e-17 & 4.256e-18 & 1.015 & \textbf{ 0.727 }\\
Output Ensemble -- PITPZ & $p_{\mathrm{out.}} (z)$ & 0.312 & 1.041e-02 & 9.166e-17 & 5.919e-18 & 1.014 & \textbf{ 1.011 }\\
\hline
\hline
   &  & \multicolumn{6}{c}{Bin 2}\\
\hline
Input Ensemble & $p_{\mathrm{in.}} (z)$ & 0.506 & 5.424e-03 & 2.007e-16 & 3.588e-18 & 1.000 & \textbf{ 1.000 }\\
Output Ensemble -- Mean Shift & - & 0.506 & 5.385e-03 & 2.013e-16 & 2.592e-18 & 1.003 & \textbf{ 0.722 }\\
Output Ensemble -- PITPZ & $p_{\mathrm{out.}} (z)$ & 0.506 & 5.387e-03 & 2.013e-16 & 3.576e-18 & 1.003 & \textbf{ 0.997 }\\
\hline
\hline
   &  & \multicolumn{6}{c}{Bin 3}\\
\hline
Input Ensemble & $p_{\mathrm{in.}} (z)$ & 0.745 & 1.998e-03 & 2.910e-16 & 6.246e-19 & 1.000 & \textbf{ 1.000 }\\
Output Ensemble -- Mean Shift & - & 0.745 & 1.999e-03 & 2.909e-16 & 4.504e-19 & 1.000 & \textbf{ 0.721 }\\
Output Ensemble -- PITPZ & $p_{\mathrm{out.}} (z)$ & 0.745 & 1.964e-03 & 2.909e-16 & 6.174e-19 & 1.000 & \textbf{ 0.989 }\\
\hline
\hline
   &  & \multicolumn{6}{c}{Bin 4}\\
\hline
Input Ensemble & $p_{\mathrm{in.}} (z)$ & 0.911 & 2.024e-03 & 3.185e-16 & 3.861e-19 & 1.000 & \textbf{ 1.000 }\\
Output Ensemble -- Mean Shift & - & 0.912 & 2.023e-03 & 3.186e-16 & 3.169e-19 & 1.000 & \textbf{ 0.821 }\\
Output Ensemble -- PITPZ & $p_{\mathrm{out.}} (z)$ & 0.912 & 2.002e-03 & 3.186e-16 & 3.830e-19 & 1.000 & \textbf{ 0.992 }\\
\hline
\hline


\end{tabular} 

\caption{Summary statistics for each ensemble in the cosmological impact analysis of this work. We show the mean value and standard deviation for each of two statistics -- mean redshift $(\overline{z})$ and lensing amplitude (\avgsigmacritinv) of an \nz, for each of four tomographic bins; we also show the relative value of the lensing amplitude mean and standard deviation compared to the input ensemble to directly highlight the difference between PITPZ and mean shifts. We find that using our PITPZ method recovers the uncertainty in $\overline{z}$ and \avgsigmacritinv of the input ensemble (the ground truth in our experiment). Using simpler mean shifts recovers only a portion of the total uncertainty in these parameters. The extent to which mean shifts underestimate uncertainty depends on the context of which underlying physical effect is being considered. In our case of photometric calibration uncertainty, we find that using mean shifts underestimates the uncertainty in lensing amplitude by as much as approximately 30 per cent in each of the bins. We choose $z=0.25$ as the lens redshift for the lensing amplitudes shown in this table.}
\label{tab:sim_like_analysis}
\end{table*}


\section{Results}
\label{sec:results}

Our primary results are shown in Figure \ref{fig:lensing_amp_conservation}, Figure \ref{fig:pit_info_conservation_cosmic_shear}, and Table \ref{tab:sim_like_analysis}. Figure \ref{fig:lensing_amp_conservation} illustrates that PITPZ propagates the relative strength of the lensing signal amplitude, which depends on higher-order moments of \nz, across all scales. By contrast, the loss of higher than mean-order moment information associated with mean shifts causes deviations from linearity in the relationship between lensing amplitude in the input ensemble and output ensemble realisations. As a result, the overall scatter in \avgsigmacritinvB is smaller in the case of using mean shifts. As shown in Table \ref{tab:sim_like_analysis}, the scatter in the output ensemble lensing amplitude using the full PITPZ method matches the true scatter from the input ensemble to within 1 per cent for all tomographic bins. By contrast, using mean shifts underestimates this scatter by 27, 28, 28, and 18 per cent in the four tomographic bins, respectively ($z_{\mathrm{lens}} = 0.25$). We can summarize the imperfections of the mean shift method relative to PITPZ in terms of two effects visually apparent in Figure \ref{fig:lensing_amp_conservation}: first, the slope of the relationship between input and output lensing amplitude deviates from the value of unity, leading to the bulk of the loss of scatter in lensing amplitude. Second, however, the mean shift method introduces significant scatter about the linear relationship, which has an overall additive effect to the scatter in the lensing amplitude. In this sense, our estimate of the degree to which mean shifts underestimate the uncertainty in lensing amplitude are a lower bound because they include this additive effect. Our result that using mean shifts on \nz underestimates uncertainty applies not only to lensing signal amplitude, but to any quantity that is a weighted integral of \nz, as any such quantity will depend on higher-order moments in \nz. We finally highlight that since \nz, unlike \avgsigmacritinv, is cosmology independent, our method does not depend on an assumed cosmology. By contrast, an attempt to propagate uncertainty by way of mean shifts on lensing signal amplitude itself would require an assumed cosmology to determine the $D_{\mathrm{ds}}$ factor present in the definition of \avgsigmacritinv. This is an additional advantage of operating directly on \nz with PITPZ. We emphasize that although the qualitative results shown are applicable in general, the quantitative difference between PITPZ and mean shifts is specific to the source of uncertainty under consideration and the redshift distributions of the source and lens galaxy samples observed. Larger values of lens redshift eliminate the impact of differences between realisations in the input ensemble at redshift values less than $z_{\mathrm{lens}}$. As one scales lens redshift up from zero, the degree to which the effect shown in Figure \ref{fig:lensing_amp_conservation} varies depends on how much relative variation in \nz is below and above the value of the lens redshift. As a result the degree to which these results change for a higher choice of lens redshift is again specific to the source of uncertainty and the redshift distribution of the galaxy survey in question. Figure \ref{fig:pit_info_conservation_cosmic_shear} shows the relationship between input and output values of the cosmic shear data vector $\xi_+$. In particular, for each $n(z)$ realisation in the input ensemble, we compute the galaxy shape two-point correlation function $\xi_+ (\theta)$ (given the assumed cosmology defined in \S \ref{sec:intro}) and the integral $\int d\theta \theta \xi_+$. We likewise compute this value for each realisation of the output ensembles produced by the mean shift and PITPZ methods, respectively. Fig. \ref{fig:pit_info_conservation_cosmic_shear} shows that PITPZ again preserves a linear relationship between input and output realisations, whereas mean shifts do not.

\section{Conclusion}
\label{sec:conclusion}

We have presented a method for transferring variations between realisations of PDFs in one ensemble onto another PDF (or ensemble of PDFs). Our method, dubbed PITPZ, may have general applications for propagating uncertainties on posterior probability functions. In addition to providing a treatment of the algorithm, we derive analytic estimates of the conservation rules for the first three moments (mean, variance, and skewness) of the PDFs used. 

We illustrate the use of this method with an experiment in the context of the weak gravitational lensing survey redshift calibration problem, for which the redshifts for large numbers of galaxies are estimated. We find that our method is an improvement over simpler mean shifts of PDFs for transferring higher-order information. We show that this higher-order information is critically important in the context of redshift calibration by propagating redshift distributions to total gravitational lensing signal amplitude, which relates directly to the cosmological constraints of lensing surveys. In summary, we find for our fiducial test case involving photometric zero-point uncertainty for a DES Y3-like survey ($z_{\mathrm{lens}} = 0.25$) that our method recovers the true uncertainty on lensing amplitude to within 1 per cent, in contrast to an underestimate of as much as 30 per cent when using mean shifts. The difference between PITPZ and mean shift on lensing amplitude reflects the importance of this method for cosmology analyses requiring redshift distributions.

We confirm that the numerical errors associated with our software implementation of our method fall well-within the LSST DESC Year 10 uncertainty targets for redshift calibration. By contrast, using simple mean shifts exceeds this uncertainty target in the mean redshift in our test case. While in our test case the error on the variance introduced by mean shifts is still so small as to fall within the LSST DESC Y10 uncertainty target in the scatter in redshift, it is the accumulated effect for all higher moments, and when also accounting for multiple independent sources of redshift uncertainty, that propagates directly to uncertainty on cosmological parameters, which may justify the additional complexity of PITPZ relative to mean shifts. Based on these results, we conclude that future galaxy lensing surveys should consider using PITPZ for propagating redshift uncertainties. 

Development of the PITPZ method has been motivated by the significant and consequential challenges of the redshift calibration problem to accomplish the stated goals of upcoming galaxy imaging surveys like the Legacy Survey of Space and Time \citep{lsst_sci_book, lsst_desc_sci_req, Ivezic2019}. In this context, improvements in our ability to measure redshift distributions from photometric galaxy samples are a prerequisite to fulfill the promise of the next generation of weak lensing experiments and of the investments made to this end. As we have discussed, PITPZ will facilitate more accurate uncertainty characterization of these measurements by enabling a transfer of uncertainties from simulations where certain observational effects can be scaled at-will to the measurements on data. Similarly, uncertainties measured in data products can be likewise transferred to measurements in simulations, which will facilitate realistic end-to-end analyses in simulations for cosmology pipeline validation. Noting the characterization of the redshift calibration problem as being within a category for which ``promising ideas exist but more exploration
is needed to determine which will work and how exactly to use them at the level
of precision needed for future surveys'' \citep{Mandelbaum2018}, we highlight that although this work has focused on weak lensing source galaxies, our method has important implications for lens redshift calibration. Given that lens redshift distributions appear as a quadratic term in the galaxy clustering signal by way of the radial selection function of lens galaxies for a given source galaxy tomographic bin (i.e. the `galaxy clustering kernel'), the galaxy clustering signal is especially sensitive to the width of the lens \nz (see e.g. \citet{y3-2x2ptbiasmodelling, y3-2x2maglimforecast, y3-galaxyclustering}). PITPZ, as a first solution to propagating \nz uncertainty for the width of \nz (and other higher than mean-order moments), may prove an essential component to calibrating lens redshift distributions within uncertainty requirements for upcoming galaxy clustering analyses. Because PITPZ is part of an effort to express redshift distribution uncertainty with sufficient complexity to meet future uncertainty goals, a natural question to ask is whether the form of redshift distribution uncertainty relates to degeneracies between redshift distribution uncertainty and other nuisance parameters in weak lensing cosmology analyses such as intrinsic alignment model parameters. We leave this question to future work. 

PITPZ is a flexible solution with numerous potential applications in the context of weak lensing redshift calibration to address the clear needs for higher precision in scheduled next-generation galaxy surveys. More broadly, recognizing the trend within astrophysics and cosmology toward the use of Bayesian statistical methods that produce full posterior probability distributions for model parameters of interest, PITPZ can serve a useful role of sophisticated propagation of uncertainties in a wide variety of sub-fields of astronomy.
\newpage
\section*{Acknowledgments}
We thank Nickolas Kokron for helpful discussion. This work was supported by the Department of Energy, Laboratory Directed Research and Development program at SLAC National Accelerator Laboratory, under contract DE-AC02-76SF00515. This work was supported by the Bavaria California Technology Center (BaCaTeC). JM acknowledges funding from the Diversifying Academia, Recruiting Excellence (DARE) and the Robert H. Siemann Fellowship Programs at Stanford University. JM thanks the LSSTC Data Science Fellowship Program, which is funded by LSSTC, NSF Cybertraining Grant \#1829740, the Brinson Foundation, and the Moore Foundation; his participation in the program has benefited this work. IH acknowledges support from the European Research Council (ERC) under the European Union’s Horizon 2020 research and innovation programme (Grant agreement No. 849169).

Funding for the DES Projects has been provided by the U.S. Department of Energy, the U.S. National Science Foundation, the Ministry of Science and Education of Spain, 
the Science and Technology Facilities Council of the United Kingdom, the Higher Education Funding Council for England, the National Center for Supercomputing 
Applications at the University of Illinois at Urbana-Champaign, the Kavli Institute of Cosmological Physics at the University of Chicago, the Center for Cosmology and Astro-Particle Physics at the Ohio State University,
the Mitchell Institute for Fundamental Physics and Astronomy at Texas A\&M University, Financiadora de Estudos e Projetos, 
Funda{\c c}{\~a}o Carlos Chagas Filho de Amparo {\`a} Pesquisa do Estado do Rio de Janeiro, Conselho Nacional de Desenvolvimento Cient{\'i}fico e Tecnol{\'o}gico and 
the Minist{\'e}rio da Ci{\^e}ncia, Tecnologia e Inova{\c c}{\~a}o, the Deutsche Forschungsgemeinschaft and the Collaborating Institutions in the Dark Energy Survey. 

The Collaborating Institutions are Argonne National Laboratory, the University of California at Santa Cruz, the University of Cambridge, Centro de Investigaciones Energ{\'e}ticas, 
Medioambientales y Tecnol{\'o}gicas-Madrid, the University of Chicago, University College London, the DES-Brazil Consortium, the University of Edinburgh, 
the Eidgen{\"o}ssische Technische Hochschule (ETH) Z{\"u}rich, 
Fermi National Accelerator Laboratory, the University of Illinois at Urbana-Champaign, the Institut de Ci{\`e}ncies de l'Espai (IEEC/CSIC), 
the Institut de F{\'i}sica d'Altes Energies, Lawrence Berkeley National Laboratory, the Ludwig-Maximilians Universit{\"a}t M{\"u}nchen and the associated Excellence Cluster Universe, 
the University of Michigan, NFS's NOIRLab, the University of Nottingham, The Ohio State University, the University of Pennsylvania, the University of Portsmouth, 
SLAC National Accelerator Laboratory, Stanford University, the University of Sussex, Texas A\&M University, and the OzDES Membership Consortium.

Based in part on observations at Cerro Tololo Inter-American Observatory at NSF's NOIRLab (NOIRLab Prop. ID 2012B-0001; PI: J. Frieman), which is managed by the Association of Universities for Research in Astronomy (AURA) under a cooperative agreement with the National Science Foundation.

The DES data management system is supported by the National Science Foundation under Grant Numbers AST-1138766 and AST-1536171.
The DES participants from Spanish institutions are partially supported by MICINN under grants ESP2017-89838, PGC2018-094773, PGC2018-102021, SEV-2016-0588, SEV-2016-0597, and MDM-2015-0509, some of which include ERDF funds from the European Union. IFAE is partially funded by the CERCA program of the Generalitat de Catalunya.
Research leading to these results has received funding from the European Research
Council under the European Union's Seventh Framework Program (FP7/2007-2013) including ERC grant agreements 240672, 291329, and 306478.
We  acknowledge support from the Brazilian Instituto Nacional de Ci\^encia
e Tecnologia (INCT) do e-Universo (CNPq grant 465376/2014-2).
\section*{Affiliations}
\label{sec:affiliations}

$^{1}$ Department of Physics, Stanford University, 382 Via Pueblo Mall, Stanford, CA 94305, USA\\
$^{2}$ Kavli Institute for Particle Astrophysics \& Cosmology, P. O. Box 2450, Stanford University, Stanford, CA 94305, USA\\
$^{3}$ SLAC National Accelerator Laboratory, Menlo Park, CA 94025, USA\\
$^{4}$ University Observatory, Faculty of Physics, Ludwig-Maximilians-Universit\"at, Scheinerstr. 1, 81679 Munich, Germany\\
$^{5}$ Institute of Astronomy, University of Cambridge, Madingley Road, Cambridge CB3 0HA, UK\\
$^{6}$ Kavli Institute for Cosmology, University of Cambridge, Madingley Road, Cambridge CB3 0HA, UK\\
$^{7}$ Argonne National Laboratory, 9700 South Cass Avenue, Lemont, IL 60439, USA\\
$^{8}$ Lawrence Berkeley National Laboratory, 1 Cyclotron Road, Berkeley, CA 94720, USA\\
$^{9}$ Jet Propulsion Laboratory, California Institute of Technology, 4800 Oak Grove Dr., Pasadena, CA 91109, USA\\
$^{10}$ Department of Physics, Carnegie Mellon University, Pittsburgh, Pennsylvania 15312, USA\\
$^{11}$ NSF AI Planning Institute for Physics of the Future, Carnegie Mellon University, Pittsburgh, PA 15213, USA\\
$^{12}$ Department of Physics and Astronomy, University of Pennsylvania, Philadelphia, PA 19104, USA\\
$^{13}$ School of Physics and Astronomy, Cardiff University, CF24 3AA, UK\\
$^{14}$ Department of Applied Mathematics and Theoretical Physics, University of Cambridge, Cambridge CB3 0WA, UK\\
$^{15}$ Department of Physics, Duke University Durham, NC 27708, USA\\
$^{16}$ Cerro Tololo Inter-American Observatory, NSF's National Optical-Infrared Astronomy Research Laboratory, Casilla 603, La Serena, Chile\\
$^{17}$ Fermi National Accelerator Laboratory, P. O. Box 500, Batavia, IL 60510, USA\\
$^{18}$ Department of Physics, University of Michigan, Ann Arbor, MI 48109, USA\\
$^{19}$ CNRS, UMR 7095, Institut d'Astrophysique de Paris, F-75014, Paris, France\\
$^{20}$ Sorbonne Universit\'es, UPMC Univ Paris 06, UMR 7095, Institut d'Astrophysique de Paris, F-75014, Paris, France\\
$^{21}$ Department of Physics \& Astronomy, University College London, Gower Street, London, WC1E 6BT, UK\\
$^{22}$ Instituto de Astrofisica de Canarias, E-38205 La Laguna, Tenerife, Spain\\
$^{23}$ Laborat\'orio Interinstitucional de e-Astronomia - LIneA, Rua Gal. Jos\'e Cristino 77, Rio de Janeiro, RJ - 20921-400, Brazil\\
$^{24}$ Universidad de La Laguna, Dpto. Astrofísica, E-38206 La Laguna, Tenerife, Spain\\
$^{25}$ Center for Astrophysical Surveys, National Center for Supercomputing Applications, 1205 West Clark St., Urbana, IL 61801, USA\\
$^{26}$ Department of Astronomy, University of Illinois at Urbana-Champaign, 1002 W. Green Street, Urbana, IL 61801, USA\\
$^{27}$ Institut de F\'{\i}sica d'Altes Energies (IFAE), The Barcelona Institute of Science and Technology, Campus UAB, 08193 Bellaterra (Barcelona) Spain\\
$^{28}$ Physics Department, William Jewell College, Liberty, MO, 64068\\
$^{29}$ Astronomy Unit, Department of Physics, University of Trieste, via Tiepolo 11, I-34131 Trieste, Italy\\
$^{30}$ INAF-Osservatorio Astronomico di Trieste, via G. B. Tiepolo 11, I-34143 Trieste, Italy\\
$^{31}$ Institute for Fundamental Physics of the Universe, Via Beirut 2, 34014 Trieste, Italy\\
$^{32}$ Hamburger Sternwarte, Universit\"{a}t Hamburg, Gojenbergsweg 112, 21029 Hamburg, Germany\\
$^{33}$ Department of Physics, IIT Hyderabad, Kandi, Telangana 502285, India\\
$^{34}$ Institute of Theoretical Astrophysics, University of Oslo. P.O. Box 1029 Blindern, NO-0315 Oslo, Norway\\
$^{35}$ Kavli Institute for Cosmological Physics, University of Chicago, Chicago, IL 60637, USA\\
$^{36}$ Instituto de Fisica Teorica UAM/CSIC, Universidad Autonoma de Madrid, 28049 Madrid, Spain\\
$^{37}$ Department of Astronomy, University of Michigan, Ann Arbor, MI 48109, USA\\
$^{38}$ Observat\'orio Nacional, Rua Gal. Jos\'e Cristino 77, Rio de Janeiro, RJ - 20921-400, Brazil\\
$^{39}$ Department of Astronomy, University of Geneva, ch. d'\'Ecogia 16, CH-1290 Versoix, Switzerland\\
$^{40}$ School of Mathematics and Physics, University of Queensland,  Brisbane, QLD 4072, Australia\\
$^{41}$ Santa Cruz Institute for Particle Physics, Santa Cruz, CA 95064, USA\\
$^{42}$ Center for Cosmology and Astro-Particle Physics, The Ohio State University, Columbus, OH 43210, USA\\
$^{43}$ Department of Physics, The Ohio State University, Columbus, OH 43210, USA\\
$^{44}$ Center for Astrophysics $\vert$ Harvard \& Smithsonian, 60 Garden Street, Cambridge, MA 02138, USA\\
$^{45}$ Australian Astronomical Optics, Macquarie University, North Ryde, NSW 2113, Australia\\
$^{46}$ Lowell Observatory, 1400 Mars Hill Rd, Flagstaff, AZ 86001, USA\\
$^{47}$ Department of Astrophysical Sciences, Princeton University, Peyton Hall, Princeton, NJ 08544, USA\\
$^{48}$ Centro de Investigaciones Energ\'eticas, Medioambientales y Tecnol\'ogicas (CIEMAT), Madrid, Spain\\
$^{49}$ Instituci\'o Catalana de Recerca i Estudis Avan\c{c}ats, E-08010 Barcelona, Spain\\
$^{50}$ Max Planck Institute for Extraterrestrial Physics, Giessenbachstrasse, 85748 Garching, Germany\\
$^{51}$ Department of Astronomy, University of California, Berkeley,  501 Campbell Hall, Berkeley, CA 94720, USA\\
$^{52}$ Department of Astronomy and Astrophysics, University of Chicago, Chicago, IL 60637, USA\\
$^{53}$ School of Physics and Astronomy, University of Southampton,  Southampton, SO17 1BJ, UK\\
$^{54}$ Computer Science and Mathematics Division, Oak Ridge National Laboratory, Oak Ridge, TN 37831\\
$^{55}$ Institute of Cosmology and Gravitation, University of Portsmouth, Portsmouth, PO1 3FX, UK\\

\section*{Data availability} 
\label{sec:release}
The DES Buzzard simulated data redshift distributions used for this work are available upon request. 



\bibliographystyle{mnras_2author}

\bibliography{references, software, y3kp}


\appendix
\onecolumn

\section{Higher order moments of Redshift Distribution}
\label{sec:conservation_appendix}

In the following we provide more complete algebra deriving the conserved quantities associated with the variance and skewness of the distributions used in our work. We use the same convention as \S \ref{sec:method}, where `in.' represents a \pz in the input ensemble that contains variation we wish to map to another \pz, $\langle \mathrm{in.} \rangle$ denotes the mean of these realisations, $T$ represents a delta transformation, and `out.' represents the output realisation resulting from the PITPZ algorithm. The index $j$ runs over the number of samples used to represent smooth \pz, as described in \S \ref{sec:implementation}.

\subsection{Variance}
We now turn to representing the variance of the output ensemble in terms of the variances of the inputs. 

\begin{equation}
    \begin{split}
        \sigma_{\mathrm{in.}}^2 &= \frac{1}{n} \sum_j^n \left[ ( z_j^{\mathrm{in.}} - \bar{z}^{\mathrm{in.}} )^2 \right] \\
        &= \frac{1}{n} \sum_j^n \left[ z_j^{\mathrm{in.}^2} - 2 z_j^{\mathrm{in.}} \bar{z}^{\mathrm{in.}} + \bar{z}^{\mathrm{in.}^2}\right].
    \end{split}
\end{equation}

\begin{equation}
\label{eqn:var_pit}
    \begin{split}
        \sigma_{T}^2 &= \frac{1}{n} \sum_j^n \left[ (\Delta_{j} - \bar{\Delta})^2 \right] \\ 
        &= \frac{1}{n} \sum_j^n \left[ (z_j^{\mathrm{in.}} - z_j^{\langle \mathrm{in.} \rangle}) - \overline{(z_j^{\mathrm{in.}} - z_j^{\langle \mathrm{in.} \rangle})} \right]^2 \\
        &= \frac{1}{n} \sum_j^n \left[ (z_j^{\mathrm{in.}} - z_j^{\langle \mathrm{in.} \rangle}) - \overline{(z_j^{\mathrm{in.}} - z_j^{\langle \mathrm{in.} \rangle})} \right] \left[ (z_j^{\mathrm{in.}} - z_j^{\langle \mathrm{in.} \rangle}) - \overline{(z_j^{\mathrm{in.}} - z_j^{\langle \mathrm{in.} \rangle})} \right] \\
        &= \frac{1}{n} \sum_j^n \left[ (z_j^{\mathrm{in.}} - z_j^{\langle \mathrm{in.} \rangle})^2 - 2 (z_j^{\mathrm{in.}} - z_j^{\langle \mathrm{in.} \rangle}) \overline{(z_j^{\mathrm{in.}} - z_j^{\langle \mathrm{in.} \rangle})} + \overline{(z_j^{\mathrm{in.}} - z_j^{\langle \mathrm{in.} \rangle})}^2 \right]\\
        &= \frac{1}{n} \sum_j^n \left[ (z_j^{\mathrm{in.}})^2 - 2 z_j^\mathrm{in.} z_j^{\langle \mathrm{in.} \rangle}  + (z_j^{\langle \mathrm{in.} \rangle})^2 - 2 z_j^{\mathrm{in.}} \overline{(z_j^\mathrm{in.} - z_j^{\langle \mathrm{in.} \rangle})} + 2 z_j^{\langle \mathrm{in.} \rangle} \overline{(z_j^\mathrm{in.} - z_j^{\langle \mathrm{in.} \rangle})} + \overline{(z_j^\mathrm{in.} - z_j^{\langle \mathrm{in.} \rangle})}^2 \right] \\
        &=  \frac{1}{n} \sum_j^n \left[ (z_j^{\mathrm{in.}})^2 - 2 z_j^\mathrm{in.} z_j^{\langle \mathrm{in.} \rangle}  + (z_j^{\langle \mathrm{in.} \rangle})^2 - 2 z_j^{\mathrm{in.}} (\bar{z}^\mathrm{in.} - \bar{z}^{\langle \mathrm{in.} \rangle}) + 2 z_j^{\langle \mathrm{in.} \rangle} (\bar{z}^\mathrm{in.} - \bar{z}^{\langle \mathrm{in.} \rangle}) + (\bar{z}^\mathrm{in.} - \bar{z}^{\langle \mathrm{in.} \rangle})^2 \right] \\
        &= \frac{1}{n} \sum_j^n \left[ (z_j^{\mathrm{in.}})^2 - 2 z_j^\mathrm{in.} z_j^{\langle \mathrm{in.} \rangle}  + (z_j^{\langle \mathrm{in.} \rangle})^2 - 2 z_j^{\mathrm{in.}} \bar{z}^\mathrm{in.} + 2 z_j^{\mathrm{in.}} \bar{z}^{\langle \mathrm{in.} \rangle} + 2 z_j^{\langle \mathrm{in.} \rangle} \bar{z}^\mathrm{in.} - 2 z_j^{\langle \mathrm{in.} \rangle} \bar{z}^{\langle \mathrm{in.} \rangle} + (\bar{z}^\mathrm{in.})^2 - 2 \bar{z}^\mathrm{in.} \bar{z}^{\langle \mathrm{in.} \rangle} + (\bar{z}^{\langle \mathrm{in.} \rangle})^2 \right] \\
        &= \frac{1}{n} \sum_j^n \left[ \left( (z_j^{\mathrm{in.}})^2 - 2 z_j^{\mathrm{in.}} \bar{z}^\mathrm{in.} + (\bar{z}^\mathrm{in.})^2 \right) + \left( (z_j^{\langle \mathrm{in.} \rangle})^2 - 2 z_j^{\langle \mathrm{in.} \rangle} \bar{z}^{\langle \mathrm{in.} \rangle} + (\bar{z}^{\langle \mathrm{in.} \rangle})^2 \right) - 2 z_j^\mathrm{in.} z_j^{\langle \mathrm{in.} \rangle}  + 2 z_j^{\mathrm{in.}} \bar{z}^{\langle \mathrm{in.} \rangle} + 2 z_j^{\langle \mathrm{in.} \rangle} \bar{z}^\mathrm{in.} - 2 \bar{z}^\mathrm{in.} \bar{z}^{\langle \mathrm{in.} \rangle} \right] \\
        &= \sigma_\mathrm{in.}^2 + \sigma_{\langle \mathrm{in.} \rangle}^2 + \frac{1}{n} \sum_j^n \left[ - 2 z_j^\mathrm{in.} z_j^{\langle \mathrm{in.} \rangle}  + 2 z_j^{\mathrm{in.}} \bar{z}^{\langle \mathrm{in.} \rangle} + 2 z_j^{\langle \mathrm{in.} \rangle} \bar{z}^\mathrm{in.} - 2 \bar{z}^\mathrm{in.} \bar{z}^{\langle \mathrm{in.} \rangle} \right] \\      &= \sigma_\mathrm{in.}^2 + \sigma_{\langle \mathrm{in.} \rangle}^2 - \frac{2}{n} \sum_j^n \left[ z_j^\mathrm{in.} z_j^{\langle \mathrm{in.} \rangle}  - z_j^{\mathrm{in.}}
        \bar{z}^{\langle \mathrm{in.} \rangle} - z_j^{\langle \mathrm{in.} \rangle} \bar{z}^\mathrm{in.} + \bar{z}^\mathrm{in.} \bar{z}^{\langle \mathrm{in.} \rangle} \right] \\    
        &= \sigma_\mathrm{in.}^2 + \sigma_{\langle \mathrm{in.} \rangle}^2 - \frac{2}{n} \sum_j^n \left[ (z_j^\mathrm{in.} - \bar{z}_j^{\mathrm{in.}}) (z_j^{\langle \mathrm{in.} \rangle} - \bar{z}^{\langle \mathrm{in.} \rangle}) \right] \\   
        &= \sigma_\mathrm{in.}^2 + \sigma_{\langle \mathrm{in.} \rangle}^2 - 2\  \mathrm{Cov} \left[ z_j^\mathrm{in.}, z_j^{\langle \mathrm{in.} \rangle} \right].
    \end{split}
\end{equation}

In summary, we find that the variance of a delta transformation can be written as the sum of the variance of the input \pz used in its construction, the variance of the mean of the realisations of the input \pz ensemble used in its construction, and the covariance between these elements. This covariance is computed directly from the ordered, evenly-spaced samples of the relevant PDFs.

\begin{equation}
    \begin{split}
        \sigma_\mathrm{out.}^2 &= \frac{1}{n} \sum_j^n \left[ \left( z_j^{\mathrm{out.}} - \bar{z}_j^{\mathrm{out.}} \right)^2 \right] \\
        &= \frac{1}{n} \sum_j^n \left[ (z_j^{\mathrm{fid.}} + \Delta_j) - \overline{(z_j^{\mathrm{fid.}} + \Delta_j)}\right]^2 \\
        &= \frac{1}{n} \sum_j^n \left[ (z_j^{\mathrm{fid.}} + \Delta_j) - \overline{(z_j^{\mathrm{fid.}} + \Delta_j)}\right] \left[ (z_j^{\mathrm{fid.}} + \Delta_j) - \overline{(z_j^{\mathrm{fid.}} + \Delta_j)}\right] \\    
        &= \frac{1}{n} \sum_j^n \left[ (z_j^{\mathrm{fid.}} + \Delta_j)^2 - 2 (z_j^{\mathrm{fid.}} + \Delta_j) \overline{(z_j^{\mathrm{fid.}} + \Delta_j)} + \overline{(z_j^{\mathrm{fid.}} + \Delta_j)}^2 \right]\\            
        &= \frac{1}{n} \sum_j^n \left[ z_j^{\mathrm{fid.}^2} + \Delta_j^2 + 2 z_j^{\mathrm{fid.}} \Delta_j - 2 z_j^{\mathrm{fid.}} (\bar{z}^{\mathrm{fid.}} + \bar{\Delta})   - 2 \Delta_j (\bar{z}^{\mathrm{fid.}} + \bar{\Delta})  + (\bar{z}^{\mathrm{fid.}} + \bar{\Delta})^2 \right]\\
        &= \frac{1}{n} \sum_j^n \left[ z_j^{\mathrm{fid.}^2} + \Delta_j^2 + 2 z_j^{\mathrm{fid.}} \Delta_j - 2 \bar{z}^{\mathrm{fid.}} z_j^{\mathrm{fid.}} - 2\bar{\Delta} z_j^{\mathrm{fid.}}  - 2 \Delta_j \bar{z}^{\mathrm{fid.}} - 2 \Delta_j \bar{\Delta} + (\bar{z}^{\mathrm{fid.}} + \bar{\Delta})^2 \right]\\
        &= \frac{1}{n} \sum_j^n \left[ z_j^{\mathrm{fid.}^2} + \Delta_j^2 + 2 z_j^{\mathrm{fid.}} \Delta_j - 2 \bar{z}^{\mathrm{fid.}} z_j^{\mathrm{fid.}} - 2\bar{\Delta} z_j^{\mathrm{fid.}}  - 2 \Delta_j \bar{z}^{\mathrm{fid.}} - 2 \Delta_j \bar{\Delta} + \bar{z}^{\mathrm{fid.}^2} + 2 \bar{z}^{\mathrm{fid.}} \bar{\Delta} + \bar{\Delta}^2 \right].\\
        &= \frac{1}{n} \sum_j^n \left[ \left( z_j^{\mathrm{fid.}^2} - 2 \bar{z}^{\mathrm{fid.}} z_j^{\mathrm{fid.}} + \bar{z}^{\mathrm{fid.}^2} \right) + \left( \Delta_j^2 - 2 \Delta_j \bar{\Delta} + \bar{\Delta}^2 \right) + 2 z_j^{\mathrm{fid.}} \Delta_j - 2\bar{\Delta} z_j^{\mathrm{fid.}}  - 2 \Delta_j \bar{z}^{\mathrm{fid.}} + 2 \bar{z}^{\mathrm{fid.}} \bar{\Delta} \right]\\
        &= \sigma_{\mathrm{fid.}}^2 + \sigma_{T}^2 + \frac{2}{n} \sum_j^n \left[ z_j^{\mathrm{fid.}} \Delta_j - \bar{\Delta} z_j^{\mathrm{fid.}}  - \Delta_j \bar{z}^{\mathrm{fid.}} + \bar{z}^{\mathrm{fid.}} \bar{\Delta} \right]\\
        &= \sigma_{\mathrm{fid.}}^2 + \sigma_{T}^2 + 2 \ \mathrm{Cov}[z_j^\mathrm{fid.}, \Delta_j].
    \end{split}
\end{equation}

Using Equation \ref{eqn:var_pit} to replace $\sigma^2_T$ with quantities from the input ensemble, this yields our final expression for the variance of a realisation in the output ensemble

\begin{equation}
    \sigma^2_\mathrm{out.} = \sigma_{\mathrm{fid.}}^2 + \sigma_\mathrm{in.}^2 + \sigma_{\langle \mathrm{in.} \rangle}^2 - 2\  \mathrm{Cov} \left[ z_j^\mathrm{in.}, z_j^{\langle \mathrm{in.} \rangle} \right] + 2 \ \mathrm{Cov}[z_j^\mathrm{fid.}, \Delta_j].
\end{equation}

Alternatively we can expand $\Delta_j$ to yield

\begin{equation}
    \sigma^2_\mathrm{out.} = \sigma_{\mathrm{fid.}}^2 + \sigma_\mathrm{in.}^2 + \sigma_{\langle \mathrm{in.} \rangle}^2 - 2\  \mathrm{Cov} \left[ z_j^\mathrm{in.}, z_j^{\langle \mathrm{in.} \rangle} \right] + 2 \ \mathrm{Cov}[z_j^\mathrm{fid.}, z_j^\mathrm{in.}] - 2 \ \mathrm{Cov} \left[ z_j^{\mathrm{fid.}}, z_j^{\langle \mathrm{in.} \rangle}\right].
\end{equation}

\subsection{Skewness}
We now turn to developing an expression for the skewness of a realisation of the output ensemble in terms of moments of the input ensemble. We use the standardized moments, which are normalized to be scale invariant. For a random variable $X$ with probability distribution $P$ with mean $\mu$, the standardized moment of degree $k$ is defined as the ratio of the moment of degree $k$ and the standard deviation $\sigma$

\begin{equation}
\widetilde{\mu}_k \equiv \frac{\mu_k}{\sigma^k} = \frac{\mathrm{E}[(X-\mu)^k]}{(\mathrm{E}[(X-\mu)^2])^{k/2}}
\end{equation}

The standardized moment of degree $k$ of a realisation of the output ensemble can be written as follows. Using $\sigma_\mathrm{out.}$ to represent the standard deviation of a given realisation (see Eqn. \ref{eqn:std_b})),

\begin{equation}
\begin{split}
\widetilde{\mu}_k^\mathrm{out.} &= \frac{1}{\sigma_\mathrm{out.}^k} \mathrm{E}[(z_j^\mathrm{out.} - \overline{z}_j^\mathrm{out.})^k]\\
&= \frac{1}{\sigma_\mathrm{out.}^k} \frac{1}{n} \sum_j^n \left[ (z_j^\mathrm{out.} - \overline{z}_j^\mathrm{out.})^k \right].
\end{split}
\end{equation}

Expanding $z_j^\mathrm{out.} = z^{\mathrm{fid.}}_j + \Delta_{j} = z^{\mathrm{fid.}}_j + z_j^\mathrm{in.} - z_j^{\langle \mathrm{in.} \rangle}$ yields

\begin{equation}
\begin{split}
\widetilde{\mu}_{k}^\mathrm{out.} &= \frac{1}{\sigma_\mathrm{out.}^k} \frac{1}{n} \sum_j^n \left[ (z^{\mathrm{fid.}}_j + z_j^\mathrm{in.} - z_j^{\langle \mathrm{in.} \rangle}) - (\overline{z^{\mathrm{fid.}}_j + z_j^\mathrm{in.} - z_j^{\langle \mathrm{in.} \rangle}}) \right]^k\\
&= \frac{1}{\sigma_\mathrm{out.}^k} \frac{1}{n} \sum_j^n \left[ (z^{\mathrm{fid.}}_j + z_j^\mathrm{in.} - z_j^{\langle \mathrm{in.} \rangle} - \overline{z}^{\mathrm{fid.}}_j - \overline{z}_j^\mathrm{in.} + \overline{z}_j^{\langle \mathrm{in.} \rangle})^k \right]\\
&= \frac{1}{\sigma_\mathrm{out.}^k} \frac{1}{n} \sum_j^n \left[ ((z^{\mathrm{fid.}}_j - \overline{z}^{\mathrm{fid.}}_j) + (z_j^\mathrm{in.} - \overline{z}_j^\mathrm{in.}) - (z_j^{\langle \mathrm{in.} \rangle} - \overline{z}_j^{\langle \mathrm{in.} \rangle}))^k \right].\\
\end{split}
\end{equation}

The standardized skewness is thus

\begin{equation}
    \begin{split}
        \widetilde{\mu}_{3}^\mathrm{out.} = \frac{1}{\sigma_\mathrm{out.}^3} \frac{1}{n} \sum_j^n [&  (z^{\mathrm{fid.}}_j - \overline{z}^{\mathrm{fid.}}_j)^3  + 3 (z^{\mathrm{fid.}}_j - \overline{z}^{\mathrm{fid.}}_j)^2 (z_j^\mathrm{in.} - \overline{z}_j^\mathrm{in.}) -3 (z^{\mathrm{fid.}}_j - \overline{z}^{\mathrm{fid.}}_j)^2 (z_j^{\langle \mathrm{in.} \rangle} - \overline{z}_j^{\langle \mathrm{in.} \rangle}) + 3 (z^{\mathrm{fid.}}_j - \overline{z}^{\mathrm{fid.}}_j) (z_j^\mathrm{in.} - \overline{z}_j^\mathrm{in.})^2 \\
        &-6 (z^{\mathrm{fid.}}_j - \overline{z}^{\mathrm{fid.}}_j) (z_j^\mathrm{in.} - \overline{z}_j^\mathrm{in.}) (z_j^{\langle \mathrm{in.} \rangle} - \overline{z}_j^{\langle \mathrm{in.} \rangle}) + 3 (z^{\mathrm{fid.}}_j - \overline{z}^{\mathrm{fid.}}_j) (z_j^{\langle \mathrm{in.} \rangle} - \overline{z}_j^{\langle \mathrm{in.} \rangle})^2 + (z_j^\mathrm{in.} - \overline{z}_j^\mathrm{in.})^3 \\
        & -3 (z_j^\mathrm{in.} - \overline{z}_j^\mathrm{in.})^2 (z_j^{\langle \mathrm{in.} \rangle} - \overline{z}_j^{\langle \mathrm{in.} \rangle}) + 3 (z_j^\mathrm{in.} - \overline{z}_j^\mathrm{in.}) (z_j^{\langle \mathrm{in.} \rangle} - \overline{z}_j^{\langle \mathrm{in.} \rangle})^2 - (z_j^{\langle \mathrm{in.} \rangle} - \overline{z}_j^{\langle \mathrm{in.} \rangle})^3].\\
    \end{split}
\end{equation}

Compare to the individual expressions for the $k^{\text{th}}$ moment of each ingredient in the recipe for constructing each realisation in the output ensemble,

\begin{equation}
\begin{split}
\widetilde{\mu}_k^\mathrm{fid.} = \frac{1}{\sigma_{\mathrm{fid.}}^k} \frac{1}{n} \sum_j^n \left[ (z_j^{\mathrm{fid.}} - \overline{z}_j^{\mathrm{fid.}})^k \right],
\end{split}
\end{equation}

\begin{equation}
\begin{split}
\widetilde{\mu}_k^\mathrm{in.} = \frac{1}{\sigma_\mathrm{in.}^k} \frac{1}{n} \sum_j^n \left[ (z_j^\mathrm{in.} - \overline{z}_j^\mathrm{in.})^k \right],
\end{split}
\end{equation}

\begin{equation}
\begin{split}
\widetilde{\mu}_k^{\langle \mathrm{in.} \rangle} = \frac{1}{\sigma_{\langle \mathrm{in.} \rangle}^k} \frac{1}{n} \sum_j^n \left[ (z_j^{\langle \mathrm{in.} \rangle} - \overline{z}_j^{\langle \mathrm{in.} \rangle})^k \right].
\end{split}
\end{equation}

We identify these terms in the expression to write the standardized skewness as

\begin{equation}
\begin{split}
    \sigma_{\mathrm{out.}}^3 \widetilde{\mu}_{3}^{\mathrm{out.}} & = \sigma_{\mathrm{fid.}}^3 \widetilde{\mu}_{3}^{\mathrm{fid.}} + \sigma_{\mathrm{in.}}^3 \widetilde{\mu}_{3}^{\mathrm{in.}} + \sigma_{\langle \mathrm{in.} \rangle}^3 \widetilde{\mu}_{3}^{\langle \mathrm{in.} \rangle}\\
    &+ 3 \sigma_{\mathrm{fid.}}^2 \sigma_{\mathrm{in.}} S(z_j^{\mathrm{fid.}}, z_j^{\mathrm{fid.}}, z_j^{\mathrm{in.}}) - 3 \sigma_{\mathrm{fid.}}^2 \sigma_{\langle \mathrm{in.} \rangle} S(z_j^{\mathrm{fid.}}, z_j^{\mathrm{fid.}}, z_j^{\langle \mathrm{in.} \rangle})\\
    &+ 3 \sigma_{\mathrm{fid.}} \sigma_{\mathrm{in.}}^2 S(z_j^{\mathrm{fid.}}, z_j^{\mathrm{in.}}, z_j^{\mathrm{in.}}) - 6 \sigma_{\mathrm{fid.}} \sigma_{\mathrm{in.}} \sigma_{\langle {\mathrm{in.}} \rangle} S(z_j^{\mathrm{fid.}}, z_j^{\mathrm{in.}}, z_j^{\langle \mathrm{in.} \rangle})\\
    &+ 3 \sigma_{\mathrm{fid.}} \sigma_{\langle \mathrm{in.} \rangle}^2 S(z_j^{\mathrm{fid.}}, z_j^{\langle \mathrm{in.} \rangle}, z_j^{\langle \mathrm{in.} \rangle}) - 3 \sigma_{\mathrm{in.}}^2 \sigma_{\langle \mathrm{in.} \rangle} S(z_j^{\mathrm{in.}}, z_j^{\mathrm{in.}}, z_j^{\langle \mathrm{in.} \rangle})\\
    &+ 3 \sigma_\mathrm{in.} \sigma_{\langle \mathrm{in.} \rangle}^2 S(z_j^{\mathrm{in.}}, z_j^{\langle \mathrm{in.} \rangle}, z_j^{\langle \mathrm{in.} \rangle}),
    \end{split}
\end{equation}

where the coskewness of three random variables $X$, $Y$, and $Z$ is defined as
\begin{equation}
S(X,Y,Z) = \frac{\text{E}[(X - \text{E}(X)) (Y - \text{E}(Y)) (Z - \text{E}(Z))]}{\sigma_X \sigma_Y \sigma_Z}.
\end{equation}

\section{Nulling}
\label{sec:nulling}

Here we introduce an additional optional procedure, which we call nulling, that can reduce the error on the mean redshift caused by the PITPZ algorithm. Nulling enforces a requirement that the mean of the delta transformation values be zero for each sample index $j$, i.e. that the mean of the delta transformations be zero for each percentile of the delta transformation distributions.

Recalling our definition of the delta transformation in \S \ref{sec:method}, we can write the $j^{\mathrm{th}}$ sample of the $i^{\mathrm{th}}$ delta transformation as the following difference in redshift values between the $i^{\mathrm{th}}$ realisation of the input ensemble ($p_i^{\mathrm{in.}}(z)$) and the mean of the input ensemble, \pzAavg.

\begin{equation}
    \begin{split}
        T_{ij} = z_{ij}^{\mathrm{in.}} - z_j^{\langle \mathrm{in.} \rangle}
    \end{split}
\end{equation}

The mean value of the $j^{\mathrm{th}}$ sample of each delta transformation over all realisations in the input ensemble is thus:

\begin{equation}
    \begin{split}
        \langle T_{ij} \rangle &= \frac{1}{n_{\mathrm{real.}}} \sum_i z_{ij}^{\mathrm{in.}} - z_j^{\langle \mathrm{in.} \rangle}\\
        &= - z_j^{\langle \mathrm{in.} \rangle} + \frac{1}{n_{\mathrm{real.}}} \sum_i z_{ij}^{\mathrm{in.}}\\
        &= - z_j^{\langle \mathrm{in.} \rangle} + \langle z_{ij}^{\mathrm{in.}} \rangle
    \end{split}
\end{equation}

This quantity does not vanish in general, in particular at the lowest and highest percentiles. These non-zero mean values at each percentile of the delta transformation sum to a non-zero mean value of the ensemble of the delta transformations.

We find empirically that without this procedure the mean of the delta transformations is approximately $10^{-5}$, which leads to an error on the mean redshift in the \pz of the output ensemble at the level of $10^{-5}$. By contrast, applying this procedure decreases the mean of the delta transformations to the level of approximately $10^{-10}$, at the expense of a slightly more complicated method and a slight deviation from the conservation rules in \S \ref{sec:conservation}.


\bsp	
\label{lastpage}

\end{document}